\documentclass{aa}

\usepackage{graphicx}
\usepackage{txfonts}
\usepackage{xcolor}
\usepackage{ulem}
\usepackage{url}
\usepackage{longtable}
\usepackage{multirow}
\usepackage{siunitx}
\usepackage{booktabs}
\usepackage{amsmath}
\usepackage{adjustbox}
\usepackage{orcidlink}
\usepackage{academicons}
\usepackage{lscape}
\usepackage{tablefootnote}
\makeatletter
\renewcommand*\aa@pageof{, page \thepage{} of \pageref*{LastPage}}

\newcommand{\kms}{\textrm{km~s}\ensuremath{^{-1}\,}}
\newcommand{\neiii}{\textrm{[Ne\,{\sc iii}]}}

\begin{document}

   \title{JWST MIRI flight performance: \\ The Medium-Resolution Spectrometer}

   \author{Ioannis~Argyriou\inst{1}\orcidlink{0000-0003-2820-1077}
          \and
          Alistair~Glasse\inst{2}\orcidlink{0000-0002-2041-2462}
          \and
          David~R.~Law\inst{3}\orcidlink{0000-0002-9402-186X}
          \and
          Alvaro~Labiano\inst{4,5}\orcidlink{0000-0002-0690-8824}
          \and
          Javier~Álvarez-Márquez\inst{5}\orcidlink{0000-0002-7093-1877}
          \and
          Polychronis~Patapis\inst{6}\orcidlink{0000-0001-8718-3732}
          \and
          Patrick~J.~Kavanagh\inst{7}\orcidlink{0000-0001-6872-2358}
          \and
          Danny~Gasman\inst{1}\orcidlink{0000-0002-1257-7742}
          \and
          Michael~Mueller\inst{8,9}
          \and
          Kirsten~Larson\inst{3}\orcidlink{0000-0003-3917-6460}
          \and
          Bart~Vandenbussche\inst{1}\orcidlink{0000-0002-1368-3109}
          \and
          Adrian~M.~Glauser\inst{6}\orcidlink{0000-0001-9250-1547}
          \and
          Pierre~Royer\inst{1}\orcidlink{0000-0001-9341-2546}
          \and
          Daniel~Dicken\inst{2}\orcidlink{0000-0003-0589-5969}
          \and
          Jake~Harkett\inst{10}
          \and
          Beth~A.~Sargent\inst{3}\orcidlink{0000-0001-9855-8261}
          \and
          Michael~Engesser\inst{3}\orcidlink{0000-0003-0209-674X}
          \and
          Olivia~C.~Jones\inst{2}\orcidlink{0000-0003-4870-5547}
          \and
          Sarah~Kendrew\inst{3,11}\orcidlink{0000-0002-7612-0469}
          \and 
          Alberto~Noriega-Crespo\inst{3}\orcidlink{0000-0002-6296-8960}
          \and 
          Bernhard~Brandl\inst{9}
          \and 
          George~H.~Rieke\inst{13}\orcidlink{0000-0003-2303-6519}
          \and 
          Gillian~S.~Wright\inst{2}\orcidlink{0000-0001-7416-7936}
          \and
          David~Lee\inst{2}
          \and
          Martyn~Wells\inst{2}
         }

   \institute{Institute of Astronomy, KU Leuven,
              Celestijnenlaan 200D, 3001 Leuven, Belgium
              \and
              UK Astronomy Technology Centre, Royal Observatory, Blackford Hill Edinburgh, EH9 3HJ, Scotland, United Kingdom
              \and 
              Space Telescope Science Institute, 3700 San Martin Drive, Baltimore, MD, 21218, USA
              \and 
              Telespazio UK for the European Space Agency, ESAC, Camino Bajo del Castillo s/n, 28692 Villanueva de la Ca\~nada, Spain \label{tpz}
              \and
              Centro de Astrobiolog\'{\i}a (CAB), CSIC-INTA, Ctra. de Ajalvir km 4, Torrej\'on de Ardoz, E-28850, Madrid, Spain \label{cab}
              \and 
              Institute of Particle Physics and Astrophysics, ETH Zürich, Wolfgang-Pauli-Str 27, 8049 Zürich Switzerland
              \and
              School of Cosmic Physics, Dublin Institute for Advanced Studies, 31 Fitzwilliam Place, Dublin 2, Ireland
              \and
              SRON Netherlands Institute for Space Research, P.O. Box 800, 9700 AV Groningen, The Netherlands
              \and
              Sterrewacht Leiden, P.O. Box 9513, 2300 RA Leiden, The Netherlands
              \and 
              Department of Physics and Astronomy, University of Leicester, University Road, Leicester, LE1 7RH, UK
              \and
              European Space Agency, 3700 San Martin Drive, Baltimore, MD, 21218, USA
              \and 
              Telophase Corporation/Code 667, NASA’s Goddard Space Flight Center, Greenbelt, MD 20771
              \and
              Steward Observatory and the Department of Astronomy, The University of Arizona, 933 N Cherry Ave, Tucson, AZ, 85750, USA\\
              \email{ioannis.argyriou@kuleuven.be}
         }

   \date{Received \today}
   
   \titlerunning{JWST MIRI MRS flight performance}

   \authorrunning{I. Argyriou \& A. Glasse \& D. Law et al.}

  \abstract
  {The Medium-Resolution Spectrometer (MRS) provides one of the four operating modes of the Mid-Infrared Instrument (MIRI) on board the James Webb Space Telescope (JWST). The MRS is an integral field spectrometer, measuring the spatial and spectral distributions of light across the 5--28~$\mu m$ wavelength range with a spectral resolving power between 3700--1300.}
   {We present the MRS’s optical, spectral, and spectro-photometric performance, as achieved in flight, and we report on the effects that limit the instrument’s ultimate sensitivity.}
   {The MRS flight performance has been quantified using observations of stars, planetary nebulae, and planets in our Solar System. The precision and accuracy of this calibration was checked against celestial calibrators with well-known flux levels and spectral features.}
   {We find that the MRS geometric calibration has a distortion solution accuracy relative to the commanded position of 8~mas at 5~$\mu m$ and 23~mas at 28~$\mu m$. The wavelength calibration is accurate to within 9~\kms at 5~$\mu m$ and 27~\kms at 28~$\mu m$. The uncertainty in the absolute spectro-photometric calibration accuracy was estimated at 5.6~$\pm$ 0.7~\%. The MIRI calibration pipeline is able to suppress the amplitude of spectral fringes to below 1.5~\% for both extended and point sources across the entire wavelength range. The MRS point spread function (PSF) is 60~\% broader than the diffraction limit along its long axis at 5~$\mu m$ and is 15~\% broader at 28~$\mu m$.}
   {The MRS flight performance is found to be better than prelaunch expectations. The MRS is one of the most subscribed observing modes of JWST and is yielding many high-profile publications. It is currently humanity’s most powerful instrument for measuring the mid-infrared spectra of celestial sources and is expected to continue as such for many years to come.}


  \maketitle
%

\section{Introduction}

\begin{figure*}[t]
\centering
\includegraphics[width=0.8\textwidth]{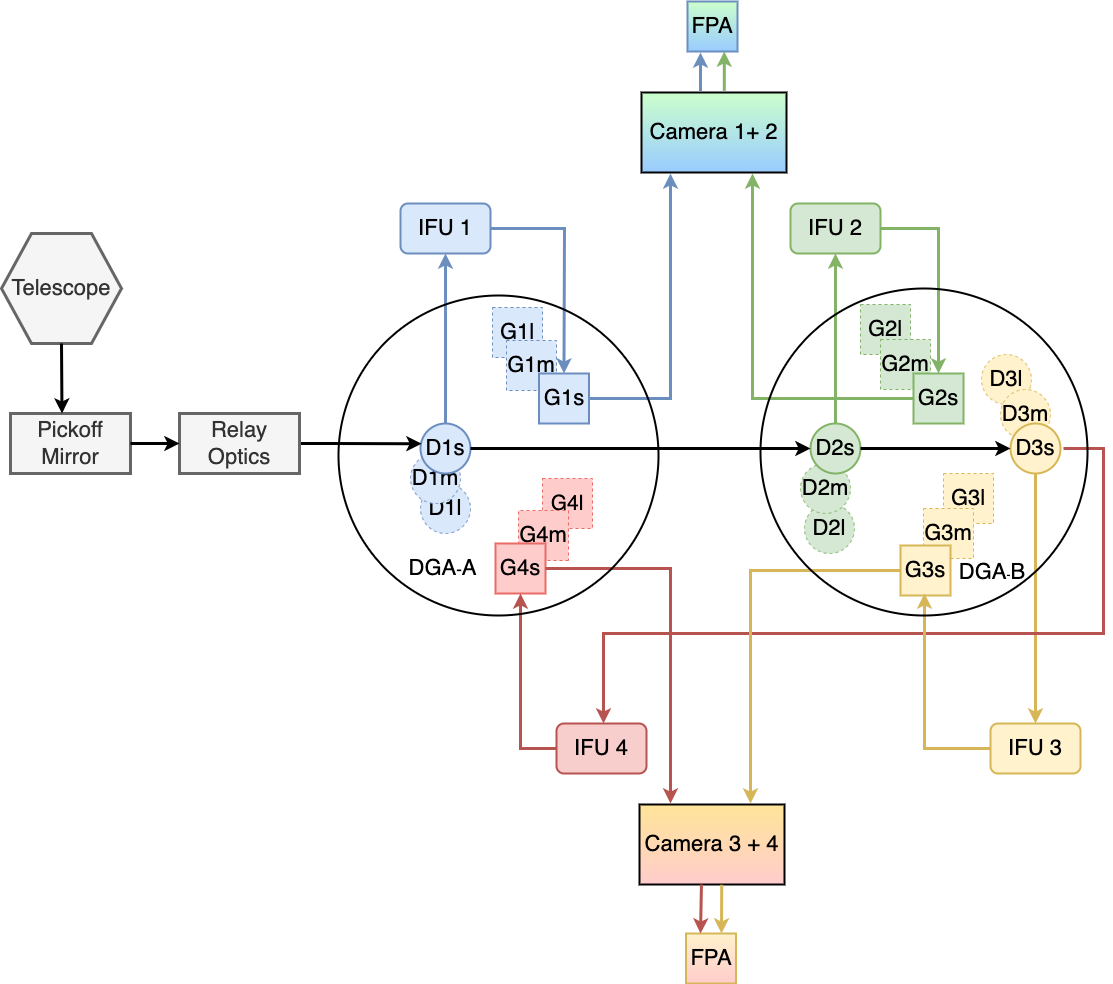}
\caption[]
{Topological layout of the MIRI MRS optics. Blue, green, orange, and red identify the instrument's four spectral channels, 1 to 4, respectively. "DGA" refers to the two dichroic-grating wheel assemblies, "IFU" to the integral field units, and "FPA" to the focal plane assemblies (the two detectors). The figure is adopted from \citet{wright23}.}
\label{fig:MRSfig}
\end{figure*}

The Mid-Infrared Instrument (MIRI), on board the \textit{James Webb Space Telescope (JWST)}, has four operating modes: (1) imaging, (2) low-resolution spectroscopy, (3) high-contrast (coronagraphic) imaging, and (4) medium-resolution spectroscopy \citep{miri_pasp_2, miri_pasp_3,miri_pasp_4,miri_pasp_5,Wells_2015}. A concise description of the operating modes can be found in \citet{wright23}, along with a summary of the in-flight performance of the respective modes. A detailed performance analysis for the MIRI Imager can be found in Dicken et al. (in prep.) and Kendrew et al. (in prep.) for the MIRI low-resolution spectrometer, and \citet{coronograph_perf} for the MIRI coronagraphic imaging. In this paper we focus on the in-flight performance of the MIRI Medium-Resolution Spectrometer \citep[MRS,][]{Wells_2015}. The high-level characterization of the overall \textit{JWST} observatory and science instrument performance from commissioning can be found in \citet{rigby23}.

The MIRI MRS is an integral-field spectrometer (IFS); in a single exposure, it measures the target's spectrum with a thousand wavelength samples and several hundred contiguous spatial samples across a rectangular field of view. For convenience, we refer to the resulting 3D dataset (two spatial and one spectral dimension) as a "data cube."

\begin{figure*}[t]
\centering
\includegraphics[width=\textwidth]{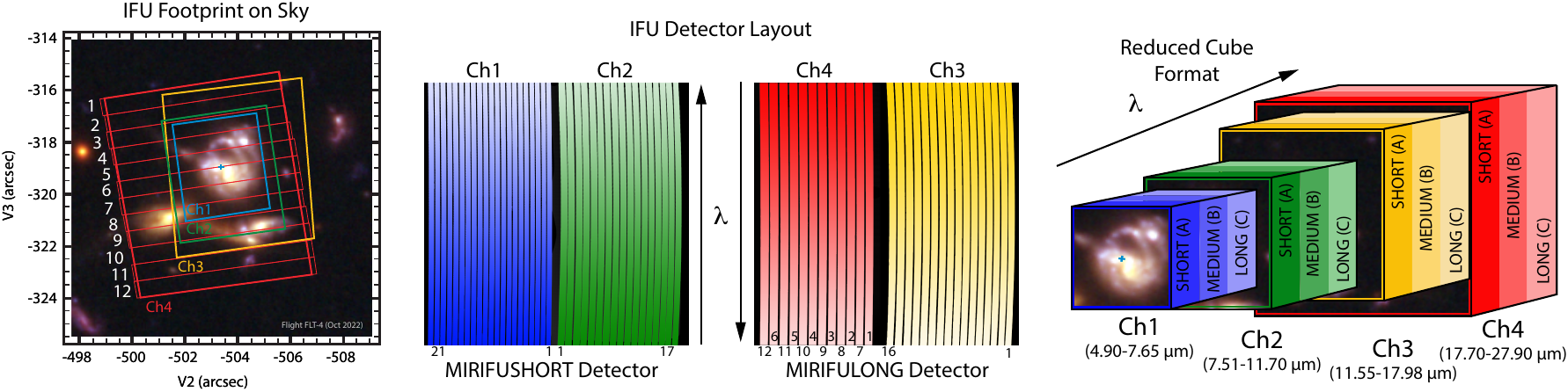}
\caption{Schematic overview of the MIRI MRS IFU.  Left-hand panel: Footprint of the IFUs for channels 1-4 (blue, green, orange, and red regions, respectively) using their "short" grating setting.  Individual slices have been drawn and numbered for band 4short. Middle panel: Layout of the dispersed spectra on the two Si:As IBC detector arrays. Slices that are adjacent on the sky are interleaved in the detector plane; numbers illustrate the schematic layout (shown in full for channel 4). Right panel: Composite data cube reconstructed from the dispersed spectra illustrating the wavelength coverage of each of the twelve MRS spectral bands. Since the IFU field of view increases with wavelength, the data cube is in practice akin to a stepped pyramid. The figure is adopted from Law et al. (accepted, ApJ).}
\label{fig:diagram}
\end{figure*}

\begin{table*}[h!]
\caption{MRS spatial and spectral sample dimensions.}
\label{tab:mrs_spectral_spatial_values}
\centering
\begin{tabular}{lcccccc}
\hline
\hline
Spectral & FOV & Nr of IFU & IFU slice & Pixel &   Wavelength & Resolving \\
Band & range [arcsec] & slices & width [arcsec] & size [arcsec] & range [$\mu m$] & power \\
\hline
1short & 3.2 x 3.7 & 21 & 0.177 & 0.196 &  4.9–5.74 & 3.320–3.710 \\
1medium & 3.2 x 3.7 & 21 & 0.177 & 0.196 & 5.66–6.63 & 3.190–3.750 \\
1long & 3.2 x 3.7 & 21 & 0.177 & 0.196 & 6.53–7.65 & 3.100–3.610 \\
2short & 4.0 x 4.8 & 17 & 0.280 & 0.196 &  7.51–8.77& 2.990–3.110 \\
2medium & 4.0 x 4.8 & 17 & 0.280 & 0.196 & 8.67–10.13 & 2.750–3.170 \\
2long & 4.0 x 4.8 & 17 & 0.280 & 0.196 & 10.02–11.70 & 2.860–3.300 \\
3short & 5.2 x 6.2 & 16 &  0.390 & 0.245 & 11.55–13.47 & 2.530–2.880 \\
3medium & 5.2 x 6.2 & 16 & 0.390 & 0.245 & 13.34–15.57 & 1.790–2.640 \\
3long & 5.2 x 6.2 & 16 & 0.390 & 0.245 & 15.41–17.98 & 1.980–2.790 \\
4short & 6.6 x 7.7 & 12 & 0.656 & 0.273 & 17.70–20.95 & 1.460–1.930 \\
4medium & 6.6 x 7.7 & 12 & 0.656 & 0.273 & 20.69–24.48 & 1.680–1.770 \\
4long & 6.6 x 7.7 & 12 & 0.656 & 0.273 & 24.19–27.9$^\alpha$ & 1.630-1.330 \\
\hline
\end{tabular}
\newline
{$^\alpha$}{This is the currently calibrated wavelength limit of the MRS. Future work may extend this limit out to 28.8~$\mu m$.}\\
\end{table*}

A block diagram showing the MRS's key optical elements is shown in Fig.~\ref{fig:MRSfig}. Light from a small (10 x 10 arcsecond) field of the telescope focal plane is selected by the pick-off mirror and presented to the two dichroic and grating wheel assemblies (DGA-A and DGA-B). In normal operations the DGAs are both set to one of three positions separated by 120 degrees, which we refer to as short, medium, and long wavelength subbands, respectively. Four spectral channels are selected by the dichroics (channels 1 to 4), with the DGA setting defining which spectral subband is selected within each channel. Referring to Fig.~\ref{fig:MRSfig}, one exposure yields bands 1short, 2short, 3short, and 4short, where each dichroic transmits wavelengths longward of (and reflects wavelengths shortward of) a fixed wavelength. Bands 1short and 2short are sliced spatially by the integral field unit (IFU) optics IFU~1 and IFU~2, and then dispersed spectrally by gratings mounted on the DGAs to rotate with their corresponding dichroic. Camera optics projected 1short and 2short onto the short wavelength (MIRIFUSHORT) focal plane assembly (FPA). Similarly, bands 3short and 4short are projected onto the long wavelength (MIRIFULONG) FPA. Both the MIRIFUSHORT and MIRIFULONG arms used the same type of Si:As impurity band conduction (IBC) array \citep{miri_pasp_7,argyriou2020SPIE}. 

The size of the MRS footprint on the sky varies with spectral channel, as shown in the left panel of Fig.~\ref{fig:diagram} in the \textit{JWST} V2/V3 coordinate system. In each channel the corresponding MRS IFU image slicer split the footprint into a distinct number of slices defined by the available detector area and the desire to match the slice width to the size of the diffraction limited PSF within each channel.  As shown in the middle panel of Fig.~\ref{fig:diagram}, each channel is imaged onto one half of the two FPAs. Importantly, slices that are adjacent on the sky are interleaved in the detector plane. The calibration effort benefited from this design decision tremendously, helping to discriminate between effects that are adjacent on the detector (e.g., dark current and scattered light) and those that are adjacent on the sky (e.g., sky background emission).

The MIRI/MRS calibration pipeline \citep{labiano2016} takes the 2D detector plane images of the dispersed spectra as input and reconstructs 3D regularized spectral cubes as shown in the right-hand panel of Fig.~\ref{fig:diagram}. These cubes have one spectral direction ($\lambda$) and two spatial directions. The cube spatial coordinates can be provided in \textit{JWST} V2/V3 coordinates, right ascension (RA), and declination (DEC), or they can be aligned to the MRS IFU local coordinates ($\alpha,\beta$). Reconstructed spectral cubes are a visually and experimentally meaningful way to analyze MRS data. However, since data cubes necessarily resample the detector plane, they can also introduce artifacts into the data (see discussion by Law et al. (accepted, ApJ)); some studies can therefore benefit from analysis of the 2d detector-level calibrated data \citep{phdthesisYannis}. In this paper we use both the cube and the detector plane level data to report on the MRS in-flight performance. The base MRS spatial and spectral sample dimensions per spectral band determined in flight are provided in Table~\ref{tab:mrs_spectral_spatial_values}.

MIRI/MRS raw data come in the form of ramps, a sequence of detector frames where the per pixel signal -- in units of data numbers (DN) -- rises with the increase in photo-charge collected as a function of time. These ramps are affected by a collection of electronic effects, explained in detail in Morrison et al. (accepted, PASP). The baseline data product assumed in this paper is a 2D MRS detector plane "slope" image, similar to the middle panel of Fig.~\ref{fig:diagram}, where the signal of each pixel has units of DN per second per pixel \citep{miri_pasp_8}. These units are converted to surface brightness units (MJy/sr) after being corrected for the instrument's spectro-photometric response and the angle subtended by the pixel on the sky.

\begin{table*}[h!]
\caption{Summary of calibration program targets. PID 1524 was a cycle 1 calibration activity that was folded into the commissioning phase. PID 1523 was a calibration activity carried out in cycle 1. PIDs 1246 and 1247 were guaranteed time observations (PI: Fletcher).\label{tab:cars}}
\centering
\begin{tabular}{cccccc}
APT PID & Targets & Target type & K-mag & Shape &  Title \\
\hline
\href{https://www.stsci.edu/jwst/science-execution/program-information.html?id=1012}{1012} & Int. Cal. Source  & Cal. Source &  - & Extended & Calibration Lamp Operation \\
\href{https://www.stsci.edu/jwst/science-execution/program-information.html?id=1024}{1024} & \href{http://simbad.cds.unistra.fr/simbad/sim-basic?Ident=MACHO+78.6706.7&submit=SIMBAD+search}{MACHO~78.6706.7}  & Star & 10.076 & Point & Plate Scale and Distortion \\
\href{https://www.stsci.edu/jwst/science-execution/program-information.html?id=1029}{1029} & \href{http://simbad.cds.unistra.fr/simbad/sim-basic?Ident=HD+37122&submit=SIMBAD+search}{HD~37122}  & Star (K2III) & 5.128 & Point & Target Acquisition Verification \\
\href{https://www.stsci.edu/jwst/science-execution/program-information.html?id=1031}{1031} & \href{http://simbad.cds.unistra.fr/simbad/sim-basic?Ident=NGC+6543&submit=SIMBAD+search}{NGC~6543}    & PNe & 8.340 & Extended & Spectral Resolution and Dispersion \\
 & \href{http://simbad.cds.unistra.fr/simbad/sim-id?Ident=HD+76534&NbIdent=1&Radius=2&Radius.unit=arcmin&submit=submit+id}{HD~76534}    & Star (Be,  B2Vn) & 7.804 & Point & " \\
 & \href{http://simbad.cds.unistra.fr/simbad/sim-id?Ident=HD+192163&NbIdent=1&Radius=2&Radius.unit=arcmin&submit=submit+id}{HD~192163}   & Star (WR) & 5.561  & Point & "  \\
\href{https://www.stsci.edu/jwst/science-execution/program-information.html?id=1039}{1039} & \href{http://simbad.cds.unistra.fr/simbad/sim-id?Ident=NGC+6552&NbIdent=1&Radius=2&Radius.unit=arcmin&submit=submit+id}{NGC~6552}    & Galaxy (Seyfert 2) & 10.323 & Extended & Detector Latent Image Characterization \\
\href{https://www.stsci.edu/jwst/science-execution/program-information.html?id=1047}{1047} & \href{http://simbad.cds.unistra.fr/simbad/sim-basic?Ident=NGC+6543&submit=SIMBAD+search}{NGC~6543}    & PNe & 8.340 & Extended & External Flat Field \\
\href{https://www.stsci.edu/jwst/science-execution/program-information.html?id=1049}{1049} & \href{http://simbad.cds.unistra.fr/simbad/sim-id?Ident=SMP+LMC+058&NbIdent=1&Radius=2&Radius.unit=arcmin&submit=submit+id}{SMP~LMC~058}    & PNe & 14.520 & Point & PSF Characterization and Spatial Resolution \\
\href{https://www.stsci.edu/jwst/science-execution/program-information.html?id=1050}{1050} & \href{http://simbad.cds.unistra.fr/simbad/sim-id?Ident=HD+158485&NbIdent=1&Radius=2&Radius.unit=arcmin&submit=submit+id}{HD~158485}  & Star (A4V) & 6.145 & Point & Photometric Sensitivity and Stability \\
 & \href{http://simbad.cds.unistra.fr/simbad/sim-id?Ident=HD+159222&NbIdent=1&Radius=2&Radius.unit=arcmin&submit=submit+id}{HD~159222}     & Star (G1V) & 4.998 & Point & " \\
 & \href{http://simbad.cds.unistra.fr/simbad/sim-id?Ident=HD+163466&NbIdent=1&Radius=2&Radius.unit=arcmin&submit=submit+id}{HD~163466}     & Star (A2) & 6.339 & Point & "\\
\href{https://www.stsci.edu/jwst/science-execution/program-information.html?id=1523}{1523} & \href{http://simbad.cds.unistra.fr/simbad/sim-basic?Ident=ngc7027&submit=SIMBAD+search}{NGC~7027} & PNe & - & Extended & External Flat Field \\
\href{https://www.stsci.edu/jwst/science-execution/program-information.html?id=1524}{1524} & \href{http://simbad.cds.unistra.fr/simbad/sim-id?Ident=10+Lac&NbIdent=1&Radius=2&Radius.unit=arcmin&submit=submit+id}{10~Lac} & Star (O9V) & 5.498 & Point & PSF Characterization \\
 \href{https://www.stsci.edu/jwst/science-execution/program-information.html?id=1246}{1246} & Jupiter & Planet & - & Extended & Jupiter's Great Red Spot \\
 \href{https://www.stsci.edu/jwst/science-execution/program-information.html?id=1246}{1247} & Saturn & Planet & - & Extended & Saturn \\
\hline
\end{tabular}
\end{table*}

\begin{table*}[h!]
\caption{Studies and analyses of MRS behavior preflight and in flight.}
\label{tab:mrs_references}
\centering
\begin{tabular}{lc}
\hline\hline
Topic & Reference\\
\hline
PhD Thesis on MIRI/MRS calibration on the ground & \citet{phdthesisYannis} \\
Wavelength calibration & \citet{labiano2021} \\
Geometric distortion and astrometric calibration & Patapis et al. (in prep.) \\
Point Spread Function & Patapis et al. (in prep.) \\
Spectral resolving power & \citet{jones23} \\
Spectral fringing & \citealt[][Mueller et al. (in prep.), Kavanagh et al. (in prep.)]{argyriou2020SPIE,argyriou2020} \\
Spectro-photometric calibration & \citet{gasman22} \\
Cube building using 3d drizzle & Law et al. (accepted, ApJ)\\
Brighter-Fatter Effect & Argyriou et al. (submitted, A\&A)\\
\hline
\end{tabular}
\end{table*}

The on-sky component of the MRS commissioning campaign took place from June 6 to June 23, 2022. In this (extremely short) amount of time, the MIRI commissioning team was able to perform all the necessary calibration activities it set out to perform and they have shown that the MRS met all of the NASA criteria necessary to be declared "ready for science," thereby closing out the commissioning phase of operations. In fact, cycle 1 calibration data and guaranteed time observation (GTO) data were used to augment commissioning data to perform the MRS performance analysis presented in this paper, where the programs used are tabulated in Table~\ref{tab:cars}. The left column provides the proposal ID numbers used in the Astronomer's Proposal Tool \href{https://www.stsci.edu/scientific-community/software/astronomers-proposal-tool-apt}{APT}. The astronomical sources selected for each calibration program shown in Table~\ref{tab:cars} were vetted prior to launch. Selection criteria included visibility at the epoch specified in the commissioning plan, the brightness and spectral shape of the source, whether these were spatially resolved, whether they had unresolved spectral lines, and whether they were photometrically variable \citep{Gordon2022}.

In Sect.~\ref{sec:calibration} we quantify the MRS in-flight performance in terms of optical fidelity, geometric and spectral distortion, and spectro-photometric precision and accuracy. We present a number of systematic effects that were not fully characterized (or even detected) on the ground. In Sect.~\ref{sec:limitations} we discuss the issues affecting the MRS's ultimate achievable sensitivity and how it compares to the preflight predictions made in \citet{miri_pasp_9}. Finally in Sect.~\ref{sec:conclusions} the conclusions of this work are presented. Throughout we provide references where appropriate to a series of other MRS technical papers that explore specific aspects of the MRS calibration in greater detail. The full list is provided in Table~\ref{tab:mrs_references}.

\section{MRS in-flight performance}
\label{sec:calibration}

\begin{figure*}
\centering
\includegraphics[width=\textwidth]{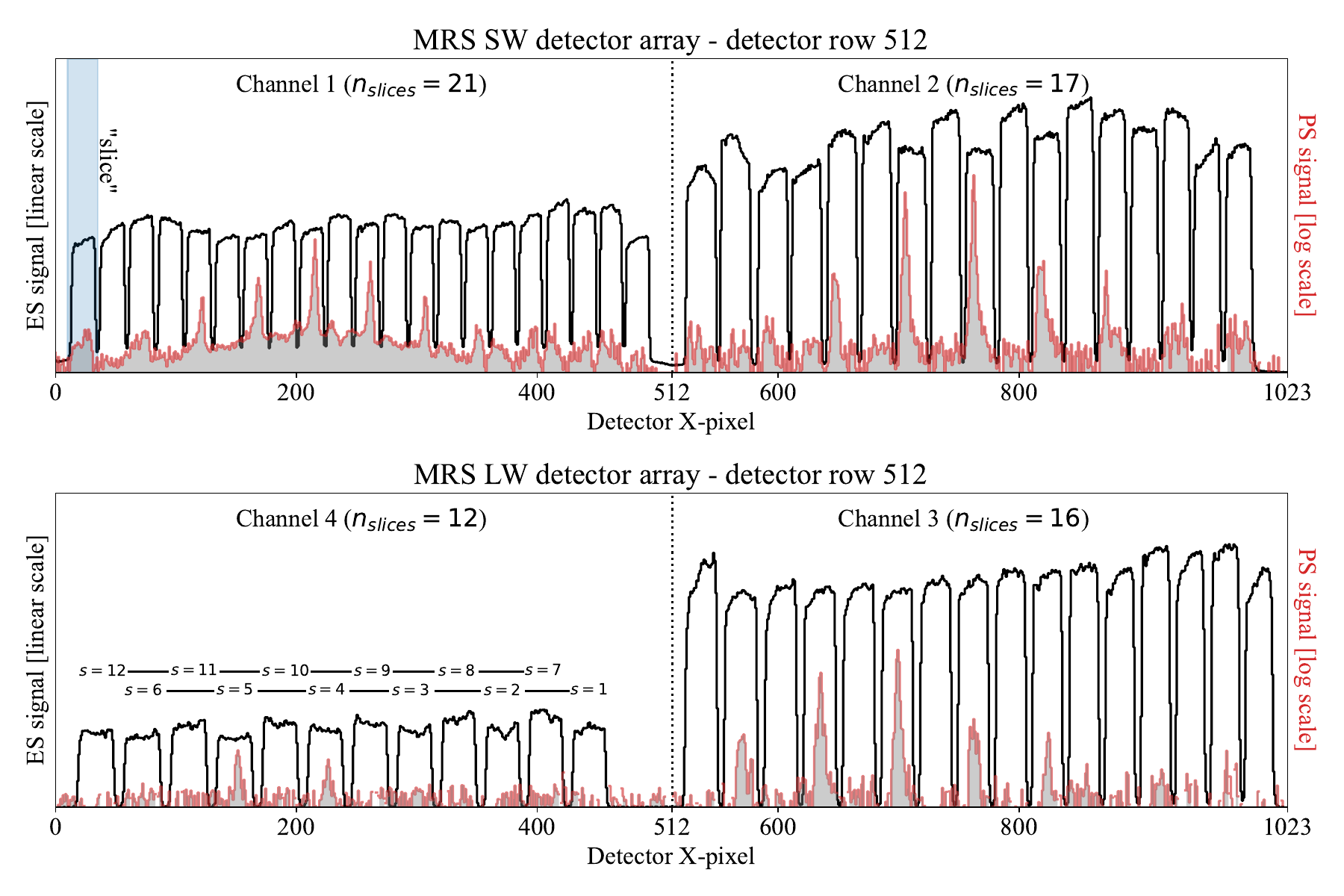}
\caption[]
{ \label{fig:detector_slice_illumination}
Extended (black, linear scale) and point (red, log scale) source illumination along a detector row.  For the point source, the telescope-plus-instrument PSF was split spatially by the integral field unit of the respective spectral channel. Different parts of the PSF were projected onto different parts of the detector in slices that do not neighbor one another on the sky. The numbering of the slices (slice number $s$) on the detector follows a similar alternating pattern for all channels to that shown for channel 4. The effect of light scattering inside the MRS detectors is clearly visible for the extended source in channels 1 and 2, where the signal between adjacent slices does not drop to zero. Although less visible in channel 2, the same effect is seen for the point source illumination.}
\end{figure*}

\subsection{Detector illumination and detector internal scattering}
\label{subsec:psf_and_scattering}

Figure~\ref{fig:detector_slice_illumination} illustrates the spatial profile of point and extended sources as they are reformatted and projected onto the MRS detectors by the MRS IFS optics. Light entering the detector experiences multiple internal reflections between the detector anti-reflection coating and the pixel metalization which cause (1) the MRS PSF to be significantly broader in the along-row direction than the diffraction limit (a phenomenon identified in the imager detector as the "cruciform" \citep{Gaspar2021}), and (2), the generation of interferometric spectral fringes, which requires careful correction in the calibration pipeline \citep{sws_fringes_and_models,fringes_sirtf_irs,stis_fringing_malumuth,argyriou2020SPIE,argyriou2020}.

\begin{figure}
\centering
\includegraphics[width=0.48\textwidth]{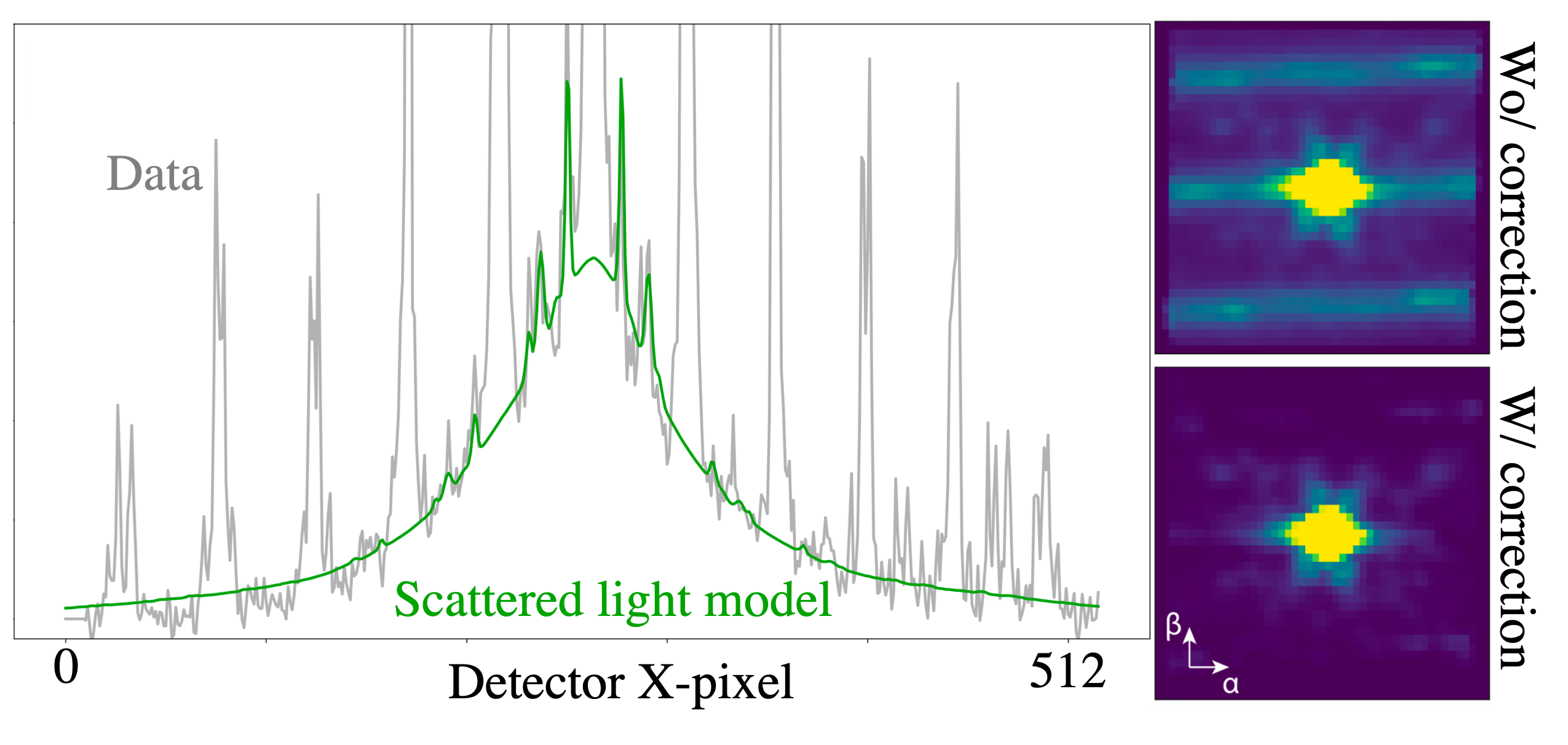}
\caption[]
{ \label{fig:scattered_light_correction}
Illustration of scattered light correction in the MIRI calibration pipeline. Left: MRS point source illumination and scattered light fitted model. Right: MRS reconstructed spectral cube image before and after scattered light correction.}
\end{figure}

The spatial behavior of the scattering is seen most clearly in the point source signal in channel 1 (Fig.~\ref{fig:scattered_light_correction}). The broad structured component is caused by diffraction of light at the narrow gaps between the MRS detector pixels \citep{miri_pasp_7,Gaspar2021}, resulting in part of the signal in one slice being detected in neighboring slices. Adjacent slices on the detector are not adjacent on the sky, so the scattered light appears in other parts of the FOV in the reconstructed 3D cube, as shown in the top right panel of Fig.~\ref{fig:scattered_light_correction}. The three diagonal stripes, one on source and two off source, are the cube manifestation of the scattered light on the detector.

The scattered light component accounts for 20~\% of the integrated power of the PSF at 5~$\mu m$, with the fraction dropping to zero at 12~$\mu m$. Due to the smaller absorption length of the arsenic-doped silicon in channels 3 and 4, scattering becomes negligible long-ward of 12~$\mu m$ with the PSF being close to diffraction limited. In order to mitigate the impact of the scattered light on the flux distribution in the reconstructed spectral cubes, an algorithm was developed to model the scattered component and subtract it directly at the detector plane level Patapis et al. (in prep.). This is shown in the left and bottom-right panels of Fig.~\ref{fig:scattered_light_correction}. The algorithm fits a series of Gaussian profiles, constrained in amplitude, to the broad and structured scattered light component, one detector row at a time, for the two spectral channels projected on the detector. The algorithm reduces the integrated power of the effect on the PSF to below the 1~\% level. This is a subtraction of flux from the 2D detector image, and comparing before and after will show a slight loss in flux.  However, the spectro-photometric calibration against standard star observations also uses the same step, so any fractional losses is made up as part of the spectro-photometric calibration step. The algorithm has been tested on semi-extended sources and single point sources, but it is recommended to examine the effectiveness of the algorithm for more complex morphologies (e.g., two resolved but blended PSFs) on a case-by-case basis.

\subsection{Geometric distortion solution}

The geometric distortion solution links the \textit{JWST} observatory V2/V3 coordinates to MRS detector pixels and vice versa. In practical terms, the location of a point source in V2/V3 may be recovered by determining the point source's centroid coordinates (pixel row, column) on the detector and applying a set of slice and wavelength specific transforms to recover the V2/V3 coordinates. 

Individual spatial and spectral distortion solutions are required for each of the MRS' 198 individual slice and spectral subband combinations  (as shown in the middle panel of Fig.~\ref{fig:diagram}) because each of the 66 slices has its own set of optics within the four IFUs and its own footprint in the common spectrometer collimator and camera optics. The accuracy of the transforms not only affects the MRS astrometry, it also impacts the fidelity of the reconstructed point source images in the 3D MRS cubes to the on-sky PSF.  Curvature of the slice illumination on the detector makes it challenging to disentangle the geometric and the spectral distortion solutions (for more information on this issue see chapter 3 of \cite{phdthesisYannis}). The methodology used to derive the MRS geometric distortion solution in each slice on the detector is described in detail in Patapis et al. (in prep.), with the accuracy achieved in each channel summarized in Table~\ref{tab:mrs_performance}.

During commissioning the preflight distortion solution (derived from the MRS Zemax optical model) was verified and updated using dithered observations of a bright star (2MASS-J05220207-6930388) in the Large Magellanic Cloud in program \href{https://www.stsci.edu/jwst/science-execution/program-information.html?id=1024}{1024}, PI: Glasse). An optimized observing program was then put together and carried out early on in Cycle 1 calibration (APT PID \href{https://www.stsci.edu/jwst/science-execution/program-information.html?id=1524}{1524}, PI: David R. Law) which observed the O-type star 10~Lac in a custom dither pattern  designed to place the unresolved source in multiple locations inside each MRS slice on the detector. 

\begin{figure}
\centering
\includegraphics[width=0.49\textwidth]{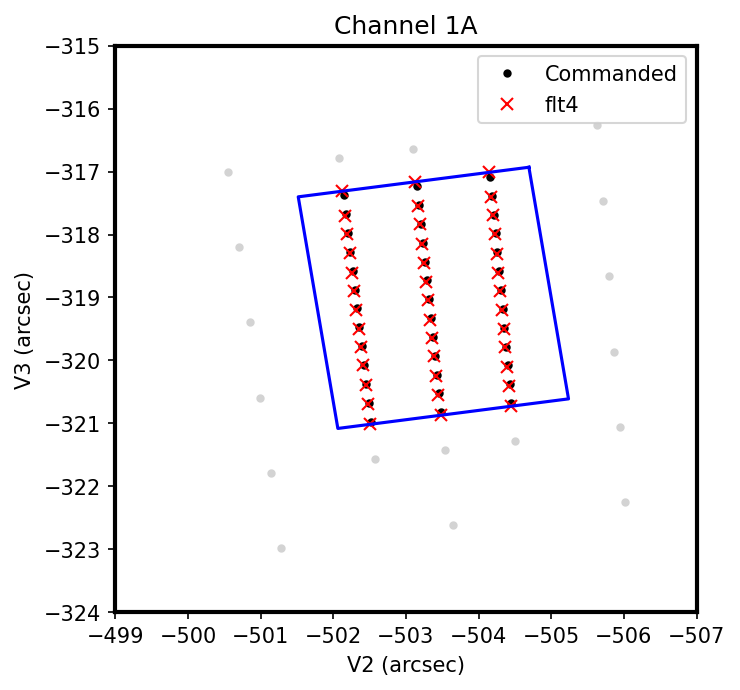}
\caption[]{Commanded versus computed pointing map of star 10~Lac in the shortest MRS spectral band (band~1short in Table~\ref{tab:mrs_spectral_spatial_values}). "flt4" represents the astrometric solution version derived and used at the start of of the MIRI Cycle 1 phase.}
\label{fig:commanded_vs_found}
\end{figure}

Fig~\ref{fig:commanded_vs_found} shows the commanded position of the star 10~Lac from PID~1524 (black dots) versus the projected positions of the star from the detector-plane-transformed V2/V3 coordinates (red crosses) for one MRS spectral band. The blue square outlines the footprint of the MRS FOV in the specific spectral band. A summary of the astrometric accuracy per band is provided in Table~\ref{tab:mrs_performance} and Patapis et al. (in prep.). We find that the astrometric accuracy is varying from 8~mas at $5~\mu m$ to 23~mas at $28~\mu m$.

\subsection{Spectral distortion solution and LSF spectral quality}

\subsubsection{Wavelength calibration}

Similar to the geometric distortion solution, the spectral distortion solution is given by a series of polynomial transformations that match each MRS detector pixel to a fixed wavelength in the rest-frame of the spacecraft. Derivation of these polynomials is nontrivial as each of the 198 slices must be calibrated individually, ideally using a reference source that is bright, fills the entire field of view, and has many narrow spectral features of known wavelength.

During the MIRI Flight Model (FM) ground test campaign, the initial wavelength calibration was performed using Fabry-Pérot etalon filters \citep{labiano2021}. The etalons provided a comb of multiple, high signal to noise ratio (S/N), unresolved spectral lines across the entire FOV in each spectral band, with absolute spectral reference features provided at a few wavelengths by a combination of wave-pass and dichroic filters. 

On orbit, there is no single ideal calibration source as the presence, position, and strength of spectral lines depends on the excitation, kinematic structure and extinction of the object observed. We have therefore adopted an iterative approach to refining the wavelength solution through successive observations of different sources. As a first-pass solution, we used observations of the bright planetary nebula NGC 6543 \citep[heliocentric radial velocity -70 \kms][]{bryce92} and the active galactic nucleus NGC 7319 \citep[$z = 0.022823$][]{hickson92} to apply a wavelength shift to the ground-based wavelength solution to match it to the strongest spectral features. The resulting "FLT-2" wavelength solution was accurate to $-27 \pm 11$ \kms\ in Ch 1A, but only to $+919 \pm 46$ \kms\ in Ch 3B.

\begin{figure*}[ht]
   \centering
    \includegraphics[width=0.98\textwidth]{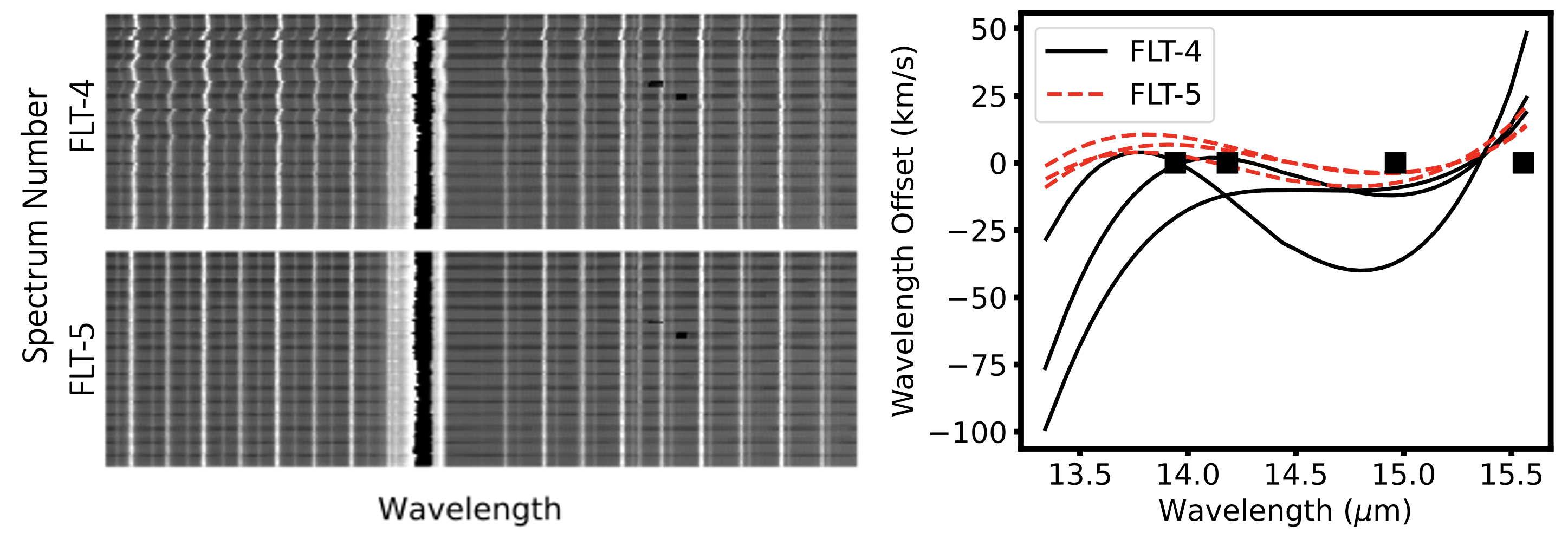}
    \caption[]{Evaluation of the wavelength distortion model and its accuracy. Left panel: Spectra of Saturn from 80 different locations throughout the Ch3B field of view plotted as a function of wavelength using the FLT-4 and FLT-5 wavelength solutions; note how FLT-5 resolves wavelength calibration artifacts seen for some locations.
    Right panel: Wavelength offset as a function of wavelength of the FLT-4 (black solid lines) and FLT-5 (red dashed lines) spectra against the atmospheric models produced using radiative transfer code NEMESIS for three arbitrarily selected positions within the Ch 3B field of view.  Solid black squares represent the locations of the HI 17-10, HI 13-9, HI 16-10, and \neiii\ $\lambda 15.5551$~$\mu m$ \ features used to constrain the FLT-4 models from NGC 6543.}
    \label{fig:wavecal}
\end{figure*}

This FLT-2 wavelength solution was nonetheless good enough to allow us to conclusively identify a host of fainter features in the spectra of NGC 6543.  Using these features, we rederived the wavelength solution based entirely on this in-flight data, fixing a number of systematic shifts (particularly in bands 3B and 3C) that had been present in the original ground wavelength solution.  This "FLT-4" wavelength solution had an accuracy of $-16 \pm 10$ \kms in Ch 1A and $-20 \pm 30$ \kms in Ch 3B, with the best performance achieved close to the wavelengths of the NGC~6543 emission features. As illustrated by Fig.~\ref{fig:wavecal} however, the FLT-4 solution could still diverge rapidly in regions of the spectrum far from these emission features where the calibration had to be extrapolated.

We therefore further revised the MRS wavelength calibration during Cycle 1 using observations of the giant planets Jupiter and Saturn (PIDs 1246+1247, PI: Fletcher).  These sources fill the MRS field of view and contain a rich forest of molecular features from 4.9 - 15~$\mu m$ \citep{fletcher16,fletcher18}.  Extracting spectra from a range of positions within each slice, we determined the wavelength offsets for bands 1A-3B by comparing to radiative transfer models produced using NEMESIS \footnote{Taking into account differential rotation of the planet across the IFU FOV.} (\citet{Irwin2008}, Harkett et al. (in prep.), Fletcher et al. (in prep.)). These offsets were used to compute a wavelength correction vector across the field of view.\footnote{We use Saturn observations
from PID 1247 to fit bands 1A, 1B, 2C-3B, and Jupiter observations from PID 1246 to fit bands 1C-2B.  These bands were
chosen according to observed data quality, saturation, and maturity of the respective atmosphere models.}
We then refit our polynomial transformations to these corrected wavelengths in order to produce the "FLT-5" wavelength solution\footnote{The FLT-5 solution is implemented in the JWST Calibration Reference Data System (CRDS) with the identifier: \href{https://jwst-crds.stsci.edu/context_table/jwst_1082.pmap}{jwst\_1082.pmap}}.

The accuracy of the FLT-5 wavelength solution was estimated by repeating the NEMESIS model comparison using the FLT-5 solution.  This comparison suggests that the latest solution is typically accurate to about 6~\kms\ (see summary in Table~\ref{tab:mrs_performance}).

As a final consistency check we compared our FLT-5 wavelength solution with observations of the bright Be star HD 76534, which contains a series of bright H recombination lines throughout the MRS wavelength range \citep[a previous version of this comparison using the FLT-4 solution was presented by][]{wright23}.  This star probes only a limited number of positions in the field of view, and at MRS spectral resolution the large rotation velocity of the star produces an asymmetric double-horned emission line profile.  We saw no evidence for significant remaining spectral shifts nor wavelength offsets between lines observed in common between adjacent wavelength
bands calibrated against Jupiter and Saturn, respectively.

\begin{figure}[h!]
\centering
\includegraphics[width=0.49\textwidth]{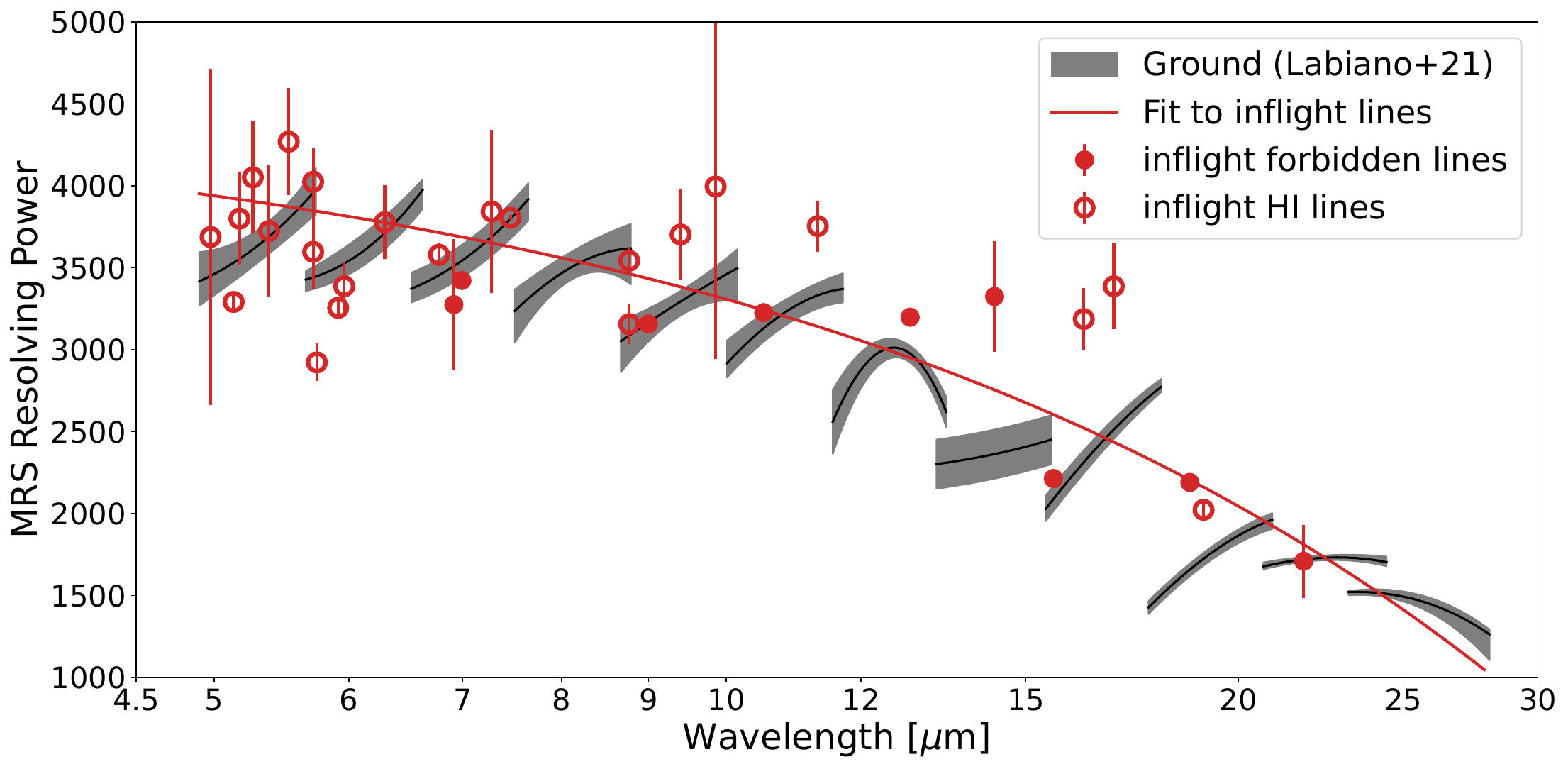}
\caption[]{MRS spectral resolution quantified using the spectrum of SMP~LMC~058 (PID~1049). Open circles are estimates using the HI recombination lines, and filled circles are estimates based on forbidden line fits. The uncertainty in the HI lines are larger due to a larger uncertainty in the two-component fit. The gray shaded areas show the MIRI ground testing estimates. A continuous line is fitted using all the datapoints.}
\label{fig:mrs_resolution}
\end{figure}

\subsubsection{Spectral resolving power}

Once the wavelength solution has been updated in all spectral bands the MRS line spread function (LSF) spectral quality can be quantified. The metric used for this is the spectral resolving power $R\sim \lambda_{cen}/FWHM$ determined from unresolved spectral lines. To compute the MRS spectral resolving power we used the emission lines in SMP~LMC~058 (APT PID~1049, Table~\ref{tab:cars}), the result is shown in Fig.~\ref{fig:mrs_resolution}. The detailed analysis of the spectral features is presented in \citet{jones23}. In summary, the detected forbidden emission lines with low ionization potential (<40eV) were found to be symmetric and unresolved. However, HI recombination lines appear to be asymmetric, with a noticeable blue tail. For the forbidden lines, a single component Gaussian fit was performed, while for the HI lines a two-component Gaussian fit was performed; the broader of the two components is adopted for the MRS resolving power. The following assumptions were used for the overall analysis: (1) that the internal velocity dispersion of SMP~LMC~058 is negligible (i.e., it cannot be detected with the MRS resolving power), (2) that the forbidden lines are unresolved, and (3) that the narrow component of the HI lines is unresolved. No high-resolution spectroscopic measurements are available to characterize the intrinsic velocity dispersion of the spectral lines of SMP~LMC~058. Nearby planetary nebulae eject gas with a typical velocity dispersion of about 10-51~km/s \citep{ReidParker2006}. Assuming a velocity dispersion of 25~km/s for SMP~LMC~058, the MRS resolving power may be underestimated by up to 5~$\%$ for channel 1, and up to 1~$\%$ for channel 4.

Overall a good match is found between the resolving power determined from the SMP~LMC~058 emission lines and the ground based estimations. As more observations of unresolved sources with unresolved spectral lines are taken with the MRS, the detailed understanding of the variation of the resolving power withing each spectral band will improve. As of now, the continuous "trend" curve in Fig.~\ref{fig:mrs_resolution} presents the state of knowledge of the MRS resolving power. The empirical relation is given by 
Eq.~\ref{eq:resolving_power}:

\begin{equation}\label{eq:resolving_power}
    R(\lambda) = 4603-128\times\lambda
.\end{equation}

\subsection{Spectral fringing}

\begin{figure}[t]
\centering
\includegraphics[width=0.49\textwidth]{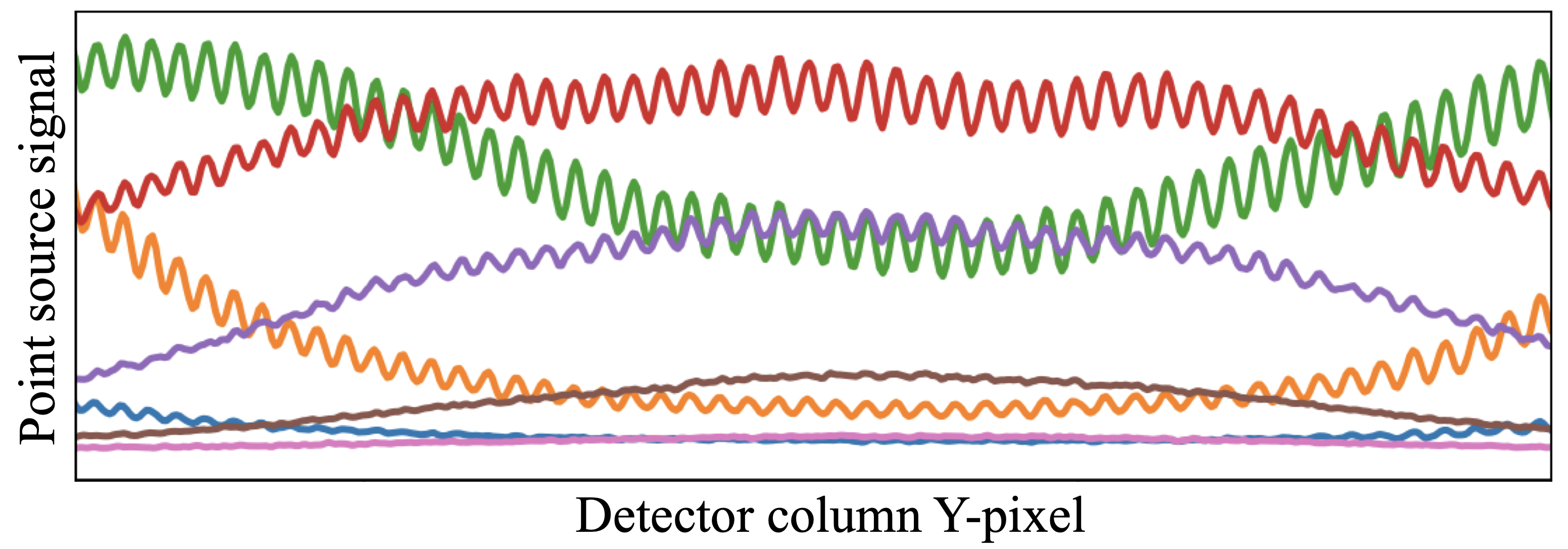}
\caption[]{Fringe pattern for different detector columns sampling the same PSF on the detector. Due to the curvature of the slices on the detector, plotting the signal in different columns results in long arcs that follow different parts of the same PSF.}
\label{fig:psf_fringe_dependency}
\end{figure}

Coherent reflections inside the MRS detectors give rise to periodic fringing in the MRS spectral dispersion direction \citep{argyriou2020SPIE,argyriou2020}. The amplitude and phase of the fringes depend on how the MIRI pupil is illuminated, namely whether the pupil is illuminated by a point source, a semi-extended source, or a fully extended source. The source extent determines the phase and amplitude of the wavefront at the detector. An example of how fringes can vary across a spectrally dispersed point source is shown in Fig.~\ref{fig:psf_fringe_dependency}. Fringes at the peak of the point source PSF are more akin to those of an extended source, where the pixel illumination is closest to being uniform in both the IFU image slicer along-slice and across-slice directions. Predicting in an analytical way how the fringe pattern of an arbitrary source depends on the incoming wavefront is a long term goal of the MRS calibration team.

Currently the MIRI pipeline corrects for the fringes using a two-step approach. First, the detector images are divided by a static fringe flat-field image derived from observations of an extended source, specifically the NGC~6543 data of PID~1031 and PID~1047 and NGC~7027 data of PID~1523. MRS fringe flats are noiseless. The uncertainty in the fringe parameters (amplitude, frequency, phase) is not provided to the pipeline. More details are provided in Mueller et al. (in prep.). Second, an empirical residual fringe correction algorithm is run on the data at the 2D detector image plane level (Kavanagh et al., in prep.). The algorithm is bootstrapped such that it only looks for frequencies that are expected based on the MRS detector layers' geometric and refractive properties. The same algorithm can be used on the 1D extracted spectrum from the spectral cubes if desired.

An alternative way to correct the fringes associated with point sources without the need for an extended source fringe flat and a residual fringe correction step is described in \citet{gasman22}, where the author shows that by using the observation of an absolute flux calibrator positioned at the exact location and pixel phase as the science target of interest, a point-source-optimized fringe flat can be derived and applied to the science target. The spectro-photometric response of the telescope and instrument can also be calibrated at the exact location and pixel phase of interest using the data and noiseless model of the absolute flux calibrator, to extract the 1D spectrum of the science target in an optimal way.

\begin{figure}[h!]
\centering
\includegraphics[width=0.49\textwidth]{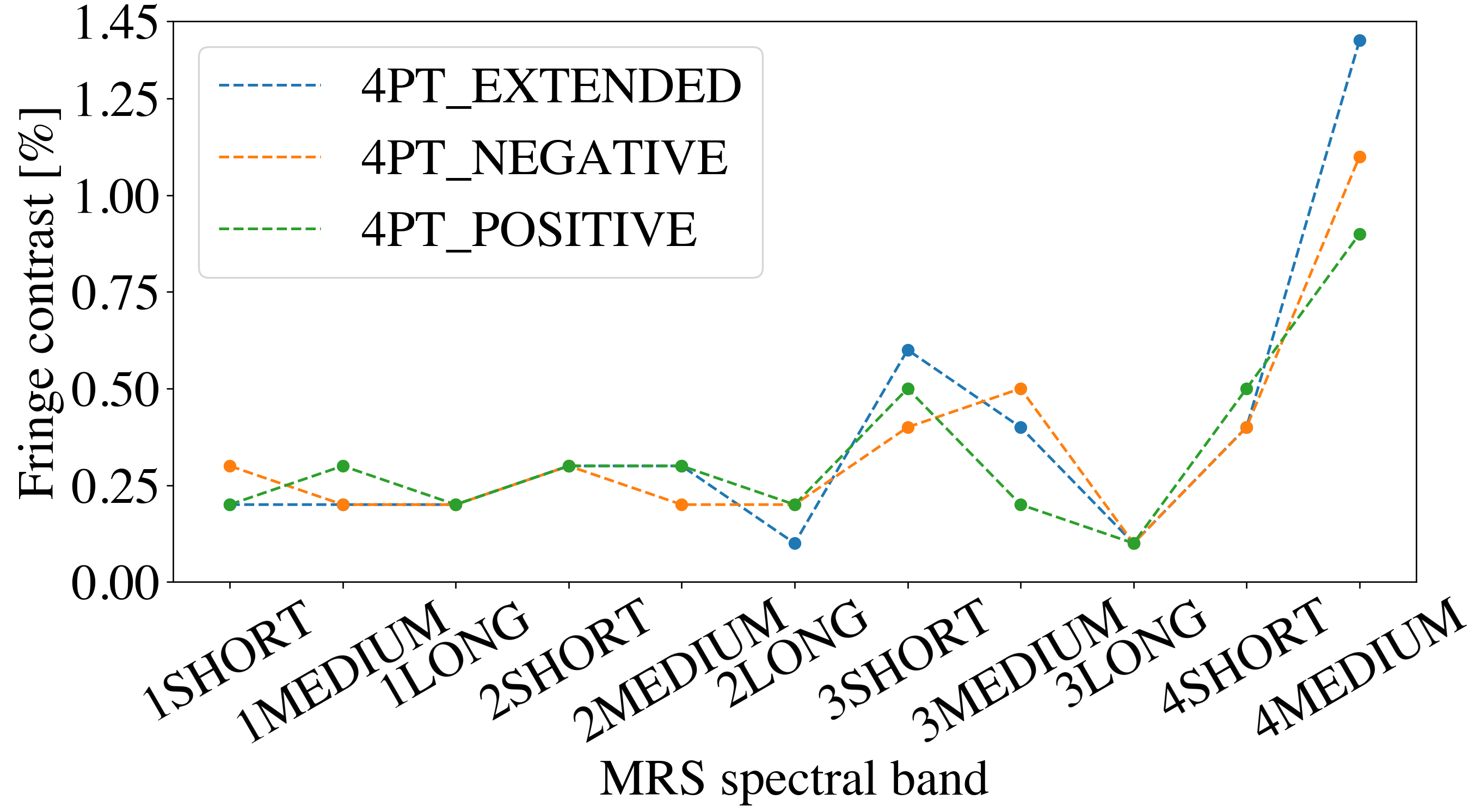}
\caption[]{Fringe statistics for all illumination types (unresolved sources, semi-extended sources) and MRS spectral bands.}
\label{fig:fringe_statistics}
\end{figure}

Fig.~\ref{fig:fringe_statistics} shows the remaining fringe contrast in the spectrum of HD~163466 (A star, PID~1050) after (1) division by the pipeline fringe flats, (2) applying the residual fringe correction algorithm on the 2D detector image, (3) applying the residual fringe correction algorithm on the 1D point source integrated spectrum extracted from the spectral cubes. The three dashed lines show three different four point dither patterns that were used to observe HD~163466. The larger contrast in bands 3short and 3medium are caused by the residuals of a beating pattern of two fringes produced in the MRS long wavelength detector \citep{argyriou2020SPIE}. The S/N in band 4long is too low to report a representative number. 

Table~\ref{tab:mrs_performance} reports the average fringe contrast in each band from the lines of Fig.~\ref{fig:fringe_statistics}. These represent the worst case scenario contrast out of all the illumination conditions (i.e., the point source case) due to the systematics linked to the nonuniform pixel illumination pattern. Critically, these numbers do not describe the impact of the residual fringe correction on the spectra of sources rich in molecular bands. This discussion is covered expensively in \citet{gasman22} where the algorithm is shown to reduce the strength of real spectral signatures. Mitigating this requires a representative fringe flat for the specific illumination pattern. As noted above, a deeper understanding of how the MIRI pupil illumination correlates with the fringe pattern on the detector is required to mitigate the need for an empirical correction. 

\subsection{MRS optical quality and aperture correction factors}
\subsubsection{Spatial resolution}

By design, the MRS is spatially and spectrally undersampled in channel 1 and channel 2 \citep{Wells_2015}, which is one of the reasons why dithered observations have always been part of the MRS' planned operational strategy. Dithering, that is to say the process of placing the point source at a different location on the detector, allows the observer to mitigate the negative impact on image quality of aliassing due to undersampling. It also allows one to reduce the impact of transient effects such as cosmic ray hits.

\begin{figure}
\centering
\includegraphics[width=0.49\textwidth]{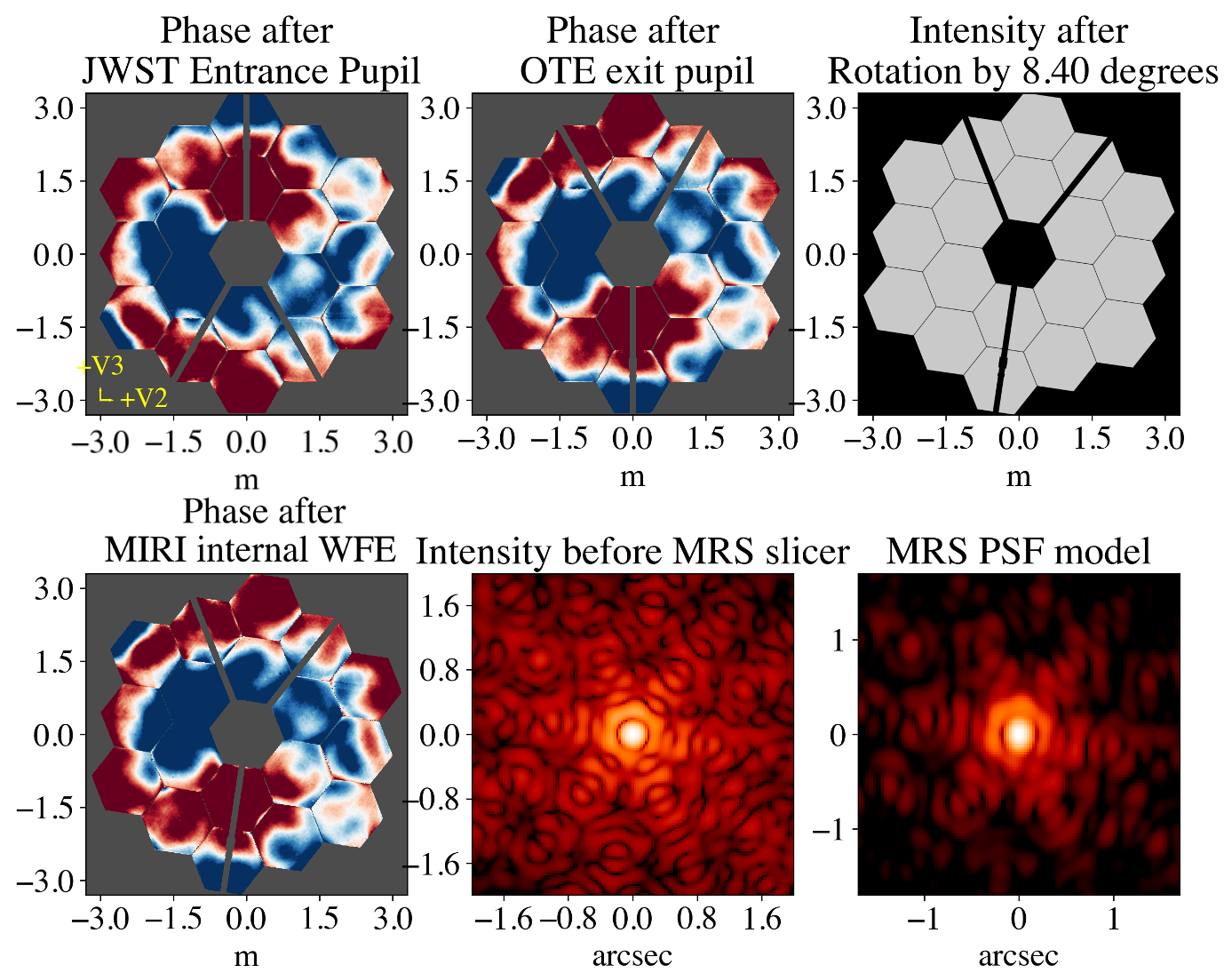}\caption[]{MIRI/MRS WebbPSF models. A model starts with the optical path difference (phase) map after the \textit{JWST} entrance pupil, which is propagated to MIRI. The MIRI internal wavefront errors are introduced, and the simulated PSF (fast Fourier transform of the phase map) is degraded to achieve the MRS PSF FWHM.}
\label{fig:mrs_webbpsf}
\end{figure}

An ideal diffraction limited model of the PSF is provided by WebbPSF \citep{Perrin2012,Perrin2014} which uses wavefront optical path difference (OPD) maps at the \textit{JWST} entrance pupil to generate wavelength-dependent MIRI PSFs. An example model at 5~$\mu m$ is shown in Fig.~\ref{fig:mrs_webbpsf}. These theoretical models form the basis of the MRS PSF characterization Patapis et al. (in prep.), where we note that the rotation of the MIRI pupil with respect to the \textit{JWST} Optical Telescope Element (OTE) is included, as shown in the top right panel. After including the MIRI plus MRS internal wavefront errors (WFEs) the intensity at the MRS image slicer and the PSF can be computed in the MRS local coordinates in arcseconds, as shown in the bottom right panel.

\begin{figure}
\centering
\includegraphics[width=0.49\textwidth]{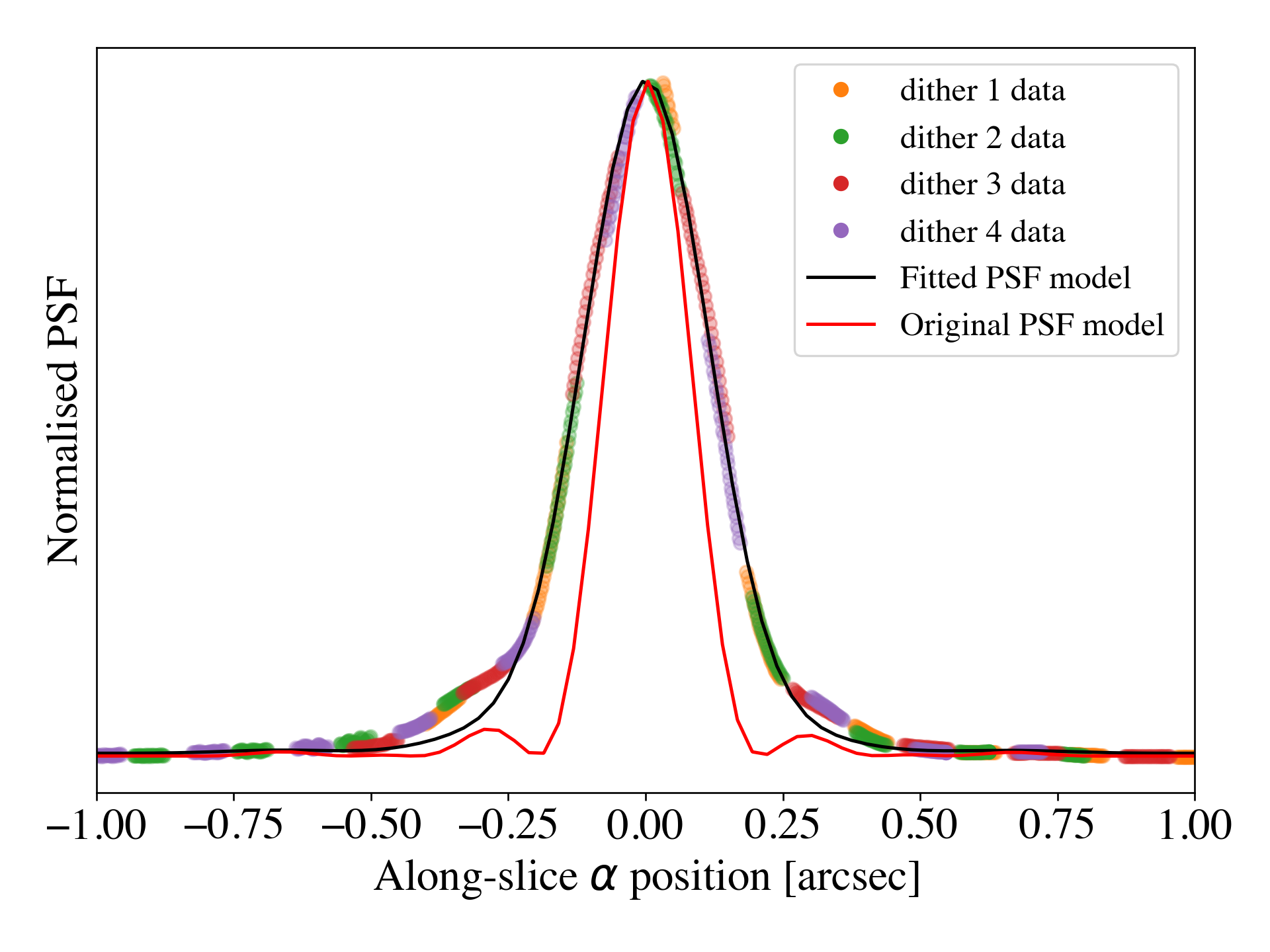}
\caption[]{MRS empirical PSF at 5.6~$\mu m$. Colored points: dither-combined commissioning observations of HD~37122 (PID~1050). Red line: Diffraction-limited WebbPSF model.  Black line: Diffraction-limited WebbPSF model (red line) convolved with a 1D Gaussian kernel.}
\label{fig:psf_fwhm_detector_fit}
\end{figure}

In Fig.~\ref{fig:psf_fwhm_detector_fit}, a 4-point dither pattern was used to observe an unresolved point source. By using (1) the spatial position of each pixel on the sky (see Patapis et al., in prep.), (2) the source centroid calculated for each dither position, and (3) the signal value in each detector row normalized by the sum of the signal in each row, separate PSF samples can be combined to make a well sampled detector PSF. In this case, we examine the core of the PSF by choosing the detector slice in the respective dithers that contains the largest integrated signal. As the metric of MRS PSF optical quality performance we use the full-width at half-maximum (FWHM) of a 1D Gaussian kernel which best fits the observed profile when convolved with the WebbPSF model.

Due to its brightness in the MRS channel 1 and 2 wavelength range, the star HD~37122 of PID~1029 was used to determined the PSF FWHM. HD~37122's brightness drops significantly toward the long wavelength bands of the MRS, leading us to use the redder spectrum of SMP~LMC~058 of PID~1049  (which is spatially unresolved) for channels 3 and 4.

\begin{figure}
\centering
\includegraphics[width=0.49\textwidth]{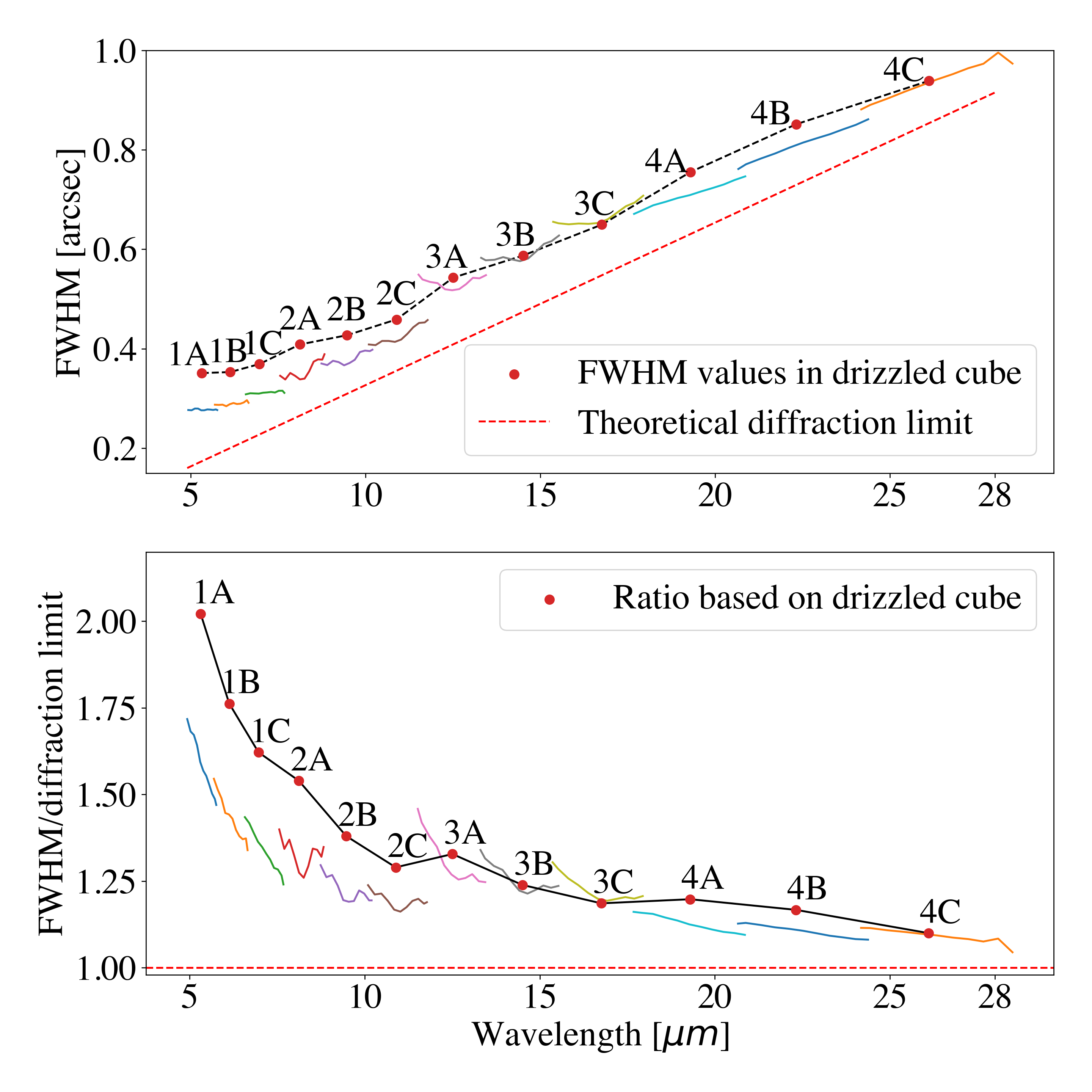}
\caption[]{MRS PSF FWHM compared to the diffraction-limited prediction. Top: MRS PSF FWHM determined on the detector image plane for each of the 12 MRS spectral bands (colored curves) as well as in the reconstructed 3D drizzled cubes (black dashed line). The theoretical diffraction limit is shown by the diagonal dashed red line. Bottom: Ratio of computed FWHM on the detector plane (colored curves) and the reconstructed 3D drizzled cubes (black dashed lines), and a diffraction-limited FWHM.}
\label{fig:mrs_detector_fwhm}
\end{figure}

In Fig.~\ref{fig:mrs_detector_fwhm} we show the values of the MRS PSF FWHM across the 12 MRS spectral bands on the detector, together with values derived from the 3D drizzled cubes (Law et al., accepted, ApJ). For the following discussion we focus on the detector FWHM values. It can clearly be seen that these are consistently above the diffraction limit at all wavelengths. For the theoretical FWHM of a diffraction limited system we use Eq.~\ref{eq:diffraction_limit}, applicable to an Airy disk\footnote{The values of Eq.~\ref{eq:diffraction_limit} are 1~\% larger than the estimated FWHM of the unconvolved WebbPSF models.}:

\begin{equation}\label{eq:diffraction_limit}
    FWHM_{diffr} = 1.025 \cdot \frac{\lambda*10^{-6}}{D_{tel}} \cdot 206264.806\, arcsec
,\end{equation}

where $D_{tel}=6.5~m$ is the diameter of the \textit{JWST} primary mirror and we used the conversion factor $1~rad = 206264.806$~arcsec.

To better visualize the difference between the computed and expected detector FWHM values, the ratio of the two is shown in the bottom panel of Fig.~\ref{fig:mrs_detector_fwhm}. From 5 to 8~$\mu m$ the ratio starts at its most significant level, and rapidly decreases. We note that the significant image broadening partially mitigates the under-sampling problem in the along-slice direction for channels 1 and 2. The broadening has been known from the MIRI ground test campaigns, however, there was never enough useful signal to accurately probe how close the PSF approached the diffraction limited past 8~$\mu m$. 

The observation of defocused PSFs in thick CCDs has previously been investigated for modern ground-based visible and near-infrared instruments \citep{lsst_defocus,moons_defocus}. The MRS' SiAs Blocked Impurity Band' (BIB) detectors are illuminated by a fast $\sim$f/3.5 focal ratio beam; with detector thicknesses of 500~$\mu m$ and 460~$\mu m$ for the short (SW) and long (LW) detectors respectively. Importantly, the size and precise speed of the MRS science beam depend on the spectral channel, the spectral band, and in a complex way on the location on the detector, having a different size in the spatial (detector x-axis) and the spectral coordinate (detector y-axis). The detector geometric thickness also varies across its surface, but for the purpose of this discussion, this variation is assumed to be negligible. 

\begin{figure}[t]
\centering
\includegraphics[width=\linewidth]{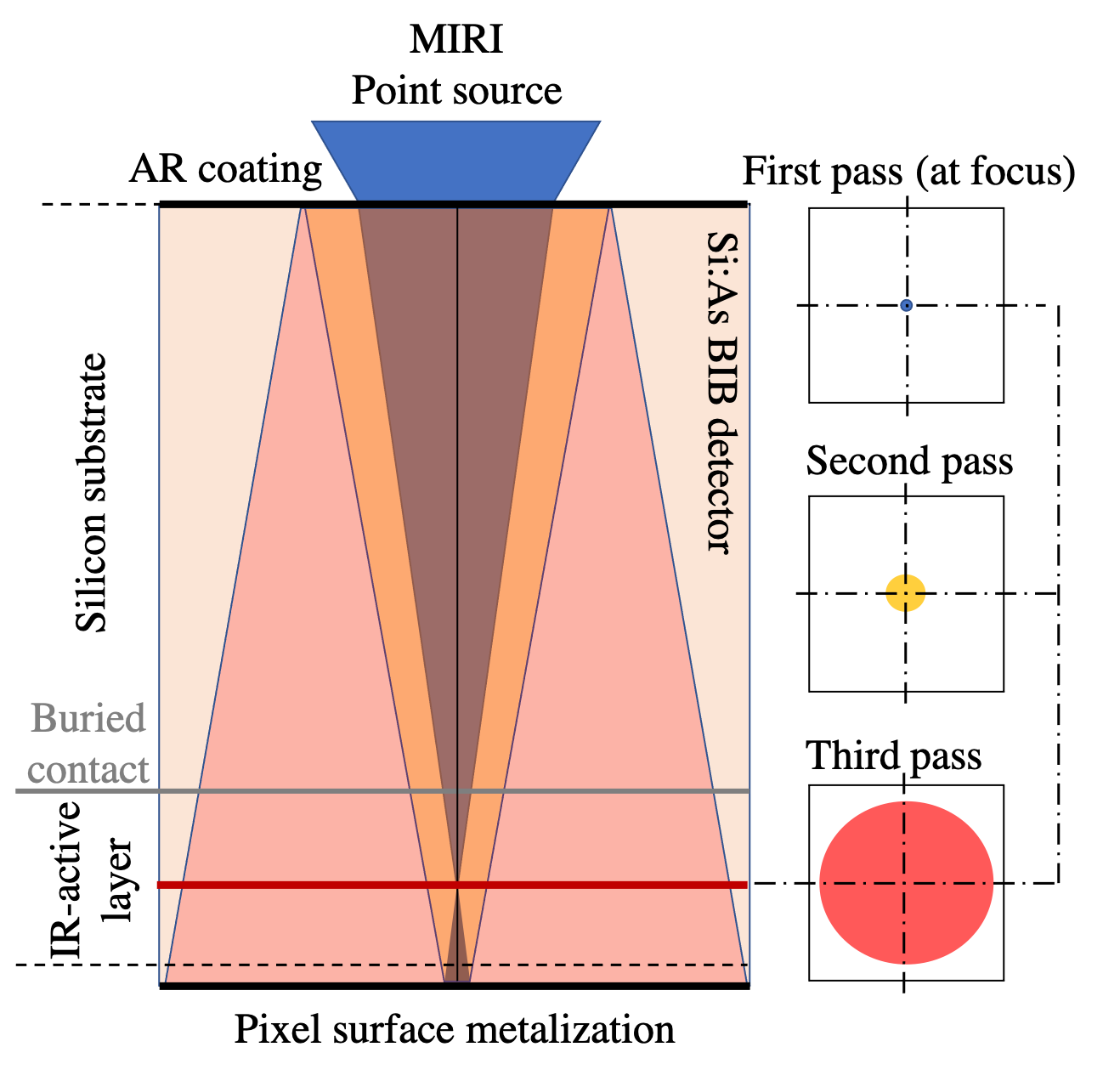}
\caption{Schematic of the MRS detector architecture (not to scale) and the process of beam divergence in the thick BIB detectors of MIRI. The final PSF is the summation of N distinct PSFs, each more defocused than the last.}
\label{fig:master_defocus}
\end{figure}

We know that the mid-infrared photons incident on the MRS detectors are reflected multiple times inside the detector structure. In Fig.~\ref{fig:master_defocus} we show the detector architecture with an illustration of the propagation of the beam from a point source. For MIRI's detectors, photon absorption happens only inside the infrared-active layer (35~$\mu m$ thick for the LW detectors, 30~$\mu m$ thick for the SW detector). Given that a majority of photons survive each pass through the infrared-active layer, this means that the MIRI/MRS PSF is the result of a summation of N distinct PSFs, where N is the number of photon passes through the infrared-active layer. This likely explains the broad profile of the oversampled PSF shown in Fig.~\ref{fig:psf_fwhm_detector_fit}, as well as the extra power in the wings of the PSF seen in Fig.~\ref{fig:psf_fwhm_detector_fit}. The jumps recorded between the MRS spectral bands and spectral channels in Fig.~\ref{fig:mrs_detector_fwhm} then correlate with the variation in the beam size between the bands and the channels as predicted by the MIRI/MRS Zemax optical model.

At wavelengths longward of 8$\mu m$, the buried contact of the arsenic-doped silicon detectors becomes increasingly reflective \citep{argyriou2020SPIE} due to the doping concentration of the thin buried contact, present between the silicon substrate and the infrared-active layer, causing a significant change in the refractive index of the layer. This picture is reinforced by the observation of spectral fringes all the way out to 28~$\mu m$. Coherent reflections then occur between the buried contact and the anti-reflection coating before entering the infrared-active layer for the first time and being absorbed. As a result, similar PSF defocussing to that which is seen at short MRS wavelengths can explain the broadening seen at longer wavelengths of the MRS as well, as shown in Fig.~\ref{fig:mrs_detector_fwhm}.

\subsubsection{Aperture correction factors}

The most straight-forward way to extract an MRS spectrum is from the 3D reconstructed spectral cubes. This can be done by defining a circular aperture radius at each wavelength and summing the signal inside that aperture. In the case of a point source observation, the PSF diffraction pattern extends much further out than the MRS FOV, hence an aperture correction factor needs to be applied to account for the part of the PSF that is outside the user-defined aperture.

The MRS aperture correction factors are determined based on the Gaussian-convolved WebbPSF models to account for (i) the large PSF size with respect to the FOV, especially in channel 4, (ii) the limited S/N of the observations, (iii) and the detector scattering making the actual PSF shape complex. The procedure to derive the aperture correction factors goes as follows. Firstly, a WebbPSF MIRI model is produced that goes out to 20 arcseconds. Secondly, the model is rotated with respect to the image axes by the same angle as the real PSF, accounting for the MRS pupil rotation (Patapis et al., in prep.). Thirdly, the model is convolved with a 1D Gaussian in the along-slice direction of the image slicer to simulate the empirically observed broadening of the PSF at each wavelength. Fourthly, the model is convolved in the image slicer across-slice direction with a 1D boxcar-function with the size of the slice width to model the finite resolution imposed by the image slicer.

\begin{figure}
\centering
\includegraphics[width=0.4\textwidth]{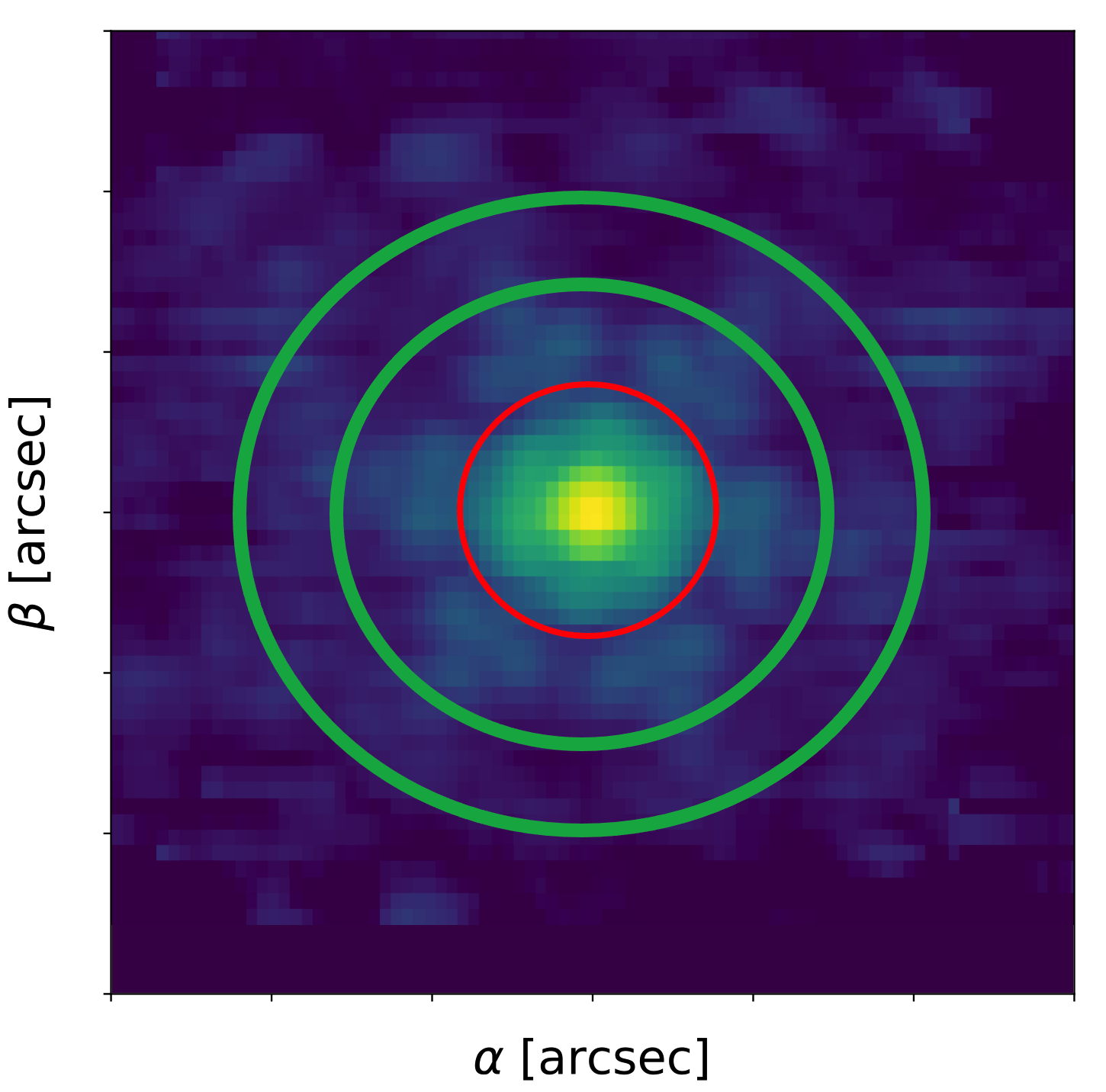}
\caption[]{Single layer of the 3D MRS reconstructed spectral cube of a point source observation. A fictitious aperture (red) is shown encircling the core of the MRS PSF. An annulus (green) is defined outside the region influenced by the MRS PSF \textit{JWST} diffraction petals in order to estimate the sky background.}
\label{fig:aperture_and_annulus}
\end{figure}

Using the set of WebbPSF models for a grid of MRS wavelengths, theoretical aperture correction factors are derived by increasing the size of a circular aperture and computing the encircled energy fraction versus the energy fraction outside the aperture. The same analysis is also run for the case where an annulus is needed to subtract the sky background from the cubes directly, illustrated in Fig.~\ref{fig:aperture_and_annulus}. Due to the MRS PSF wings extending out to the full MRS FOV (see bottom right panel of Fig.~\ref{fig:mrs_webbpsf}), we reintroduce the part of the PSF included in the annulus in the aperture correction factors.

\begin{figure}
\centering
\includegraphics[width=0.49\textwidth]{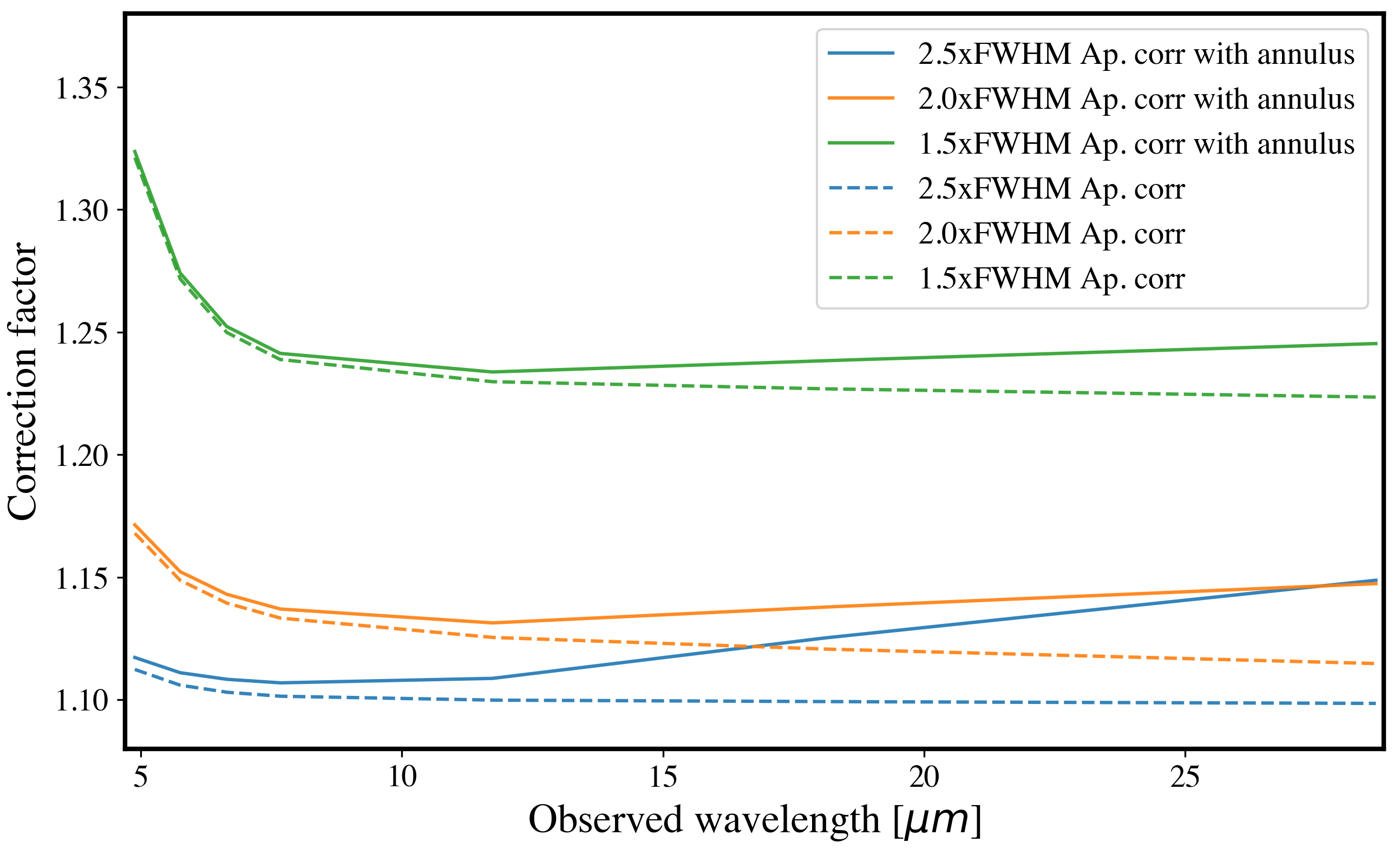}
\caption[]{MRS aperture correction factors with and without annulus for background subtraction.}
\label{fig:apercorr_factors}
\end{figure}

Figure~\ref{fig:apercorr_factors} shows the MRS aperture correction factors as a function of wavelength, assuming a circular aperture with a radius that grows with the measured PSF FWHM. Detector scattering results in an initial down-tick in the correction factors from 5 to 10~$\mu m$, after which the values gradually rise. For faint sources a smaller aperture radius is needed in order to minimize the contribution of the background signal. Similarly, the size of the aperture is limited due to the small MRS FOV. In the MRS spectral channel 4, at the longest wavelengths, the MRS PSF petals cover the entirety of the FOV\footnote{Based on the MRS PSF model linked to the CRDS version jwst\_1082.pmap, for the default aperture radius of 2.5$\times$FWHM, at the short wavelength end, 92.5\% of the PSF is contained inside the aperture and $\sim$1\% of the PSF is contained in the annulus. At the long wavelength end, 91\% of the PSF is contained in the aperture and $\sim$3\%  of the PSF is contained in the corresponding annulus. By spectro-photometrically calibrating the MRS using a point-source and the same aperture and annulus sizes, we make sure that this systematic effect is accounted for.}. Dedicated "off-source" background observations can be used to mitigate this issue.

\subsection{Spectro-photometric response}

The precise spectro-photometric response as a function of wavelength is needed to accurately predict the spectral flux density from a source, regardless of whether it is resolved or unresolved, regardless of its spectral color, and regardless of its position in the MRS FOV. The reality is that the wavelength-dependent detector pixel nonlinear response, the behavior of the spectral fringes, the estimation of the radiative (sky) background, the uncertainties in theoretical stellar models and aperture correction factors, and the time evolution of the end-to-end system response can make the task of precisely determining the wavelength relative (and absolute) spectral flux density of a source challenging. We describe the approach taken for the initial spectro-photometric solution based on commissioning observations. This work will be superseded by on-going efforts to rederive a fully flight-based calibration using a variety of standard stars observed throughout the Cycle 1 calibration program. This Cycle 1 calibration program will assess and account for any evolution in photometric sensitivity that may be observed during the first year of observations.

The MIRI MRS spectro-photometric response was determined on the ground using a spatially uniform extended blackbody source (see chapter 7 in \cite{phdthesisYannis} for more details). Unfortunately in the mid-infrared there are no sufficiently spatially uniform extended sources in the sky. For that reason, the MRS calibration strategy during commissioning was to update the 2D spectro-photometric response on the detector using the 1D spectrum of the A-star HD~163466, flux calibrated using a theoretical continuum model \citep{Gordon2022}.

The process of rederiving the MRS spectro-photometric response then involved the following steps. Firstly, we reduce the A-star HD~163466 observations in the 12 MRS spectral bands using the ground-based spectro-photometric calibration and create 3D spectral cubes. In this case a dedicated background observation was used to subtract the sky background. This subtraction was performed using the 2D detector images. Secondly, we extract the 1D integrated spectrum of HD~163466 using an aperture in the spectral cubes and apply the corresponding aperture correction factors. Thirdly, we fit the spectral continuum of HD~163466 with a third order spline, making sure only continuum values are used for the spline knot points. A theoretical model is used to pinpoint the location of spectral lines and avoid them for the continuum definition \citep{Bohlin2014}. Fourthly, we divide the fitted continuum by the theoretical continuum of HD~163466. This  correction is then applied to the ground-based spectro-photometric solution.

\begin{figure}
\centering
\includegraphics[width=0.49\textwidth]{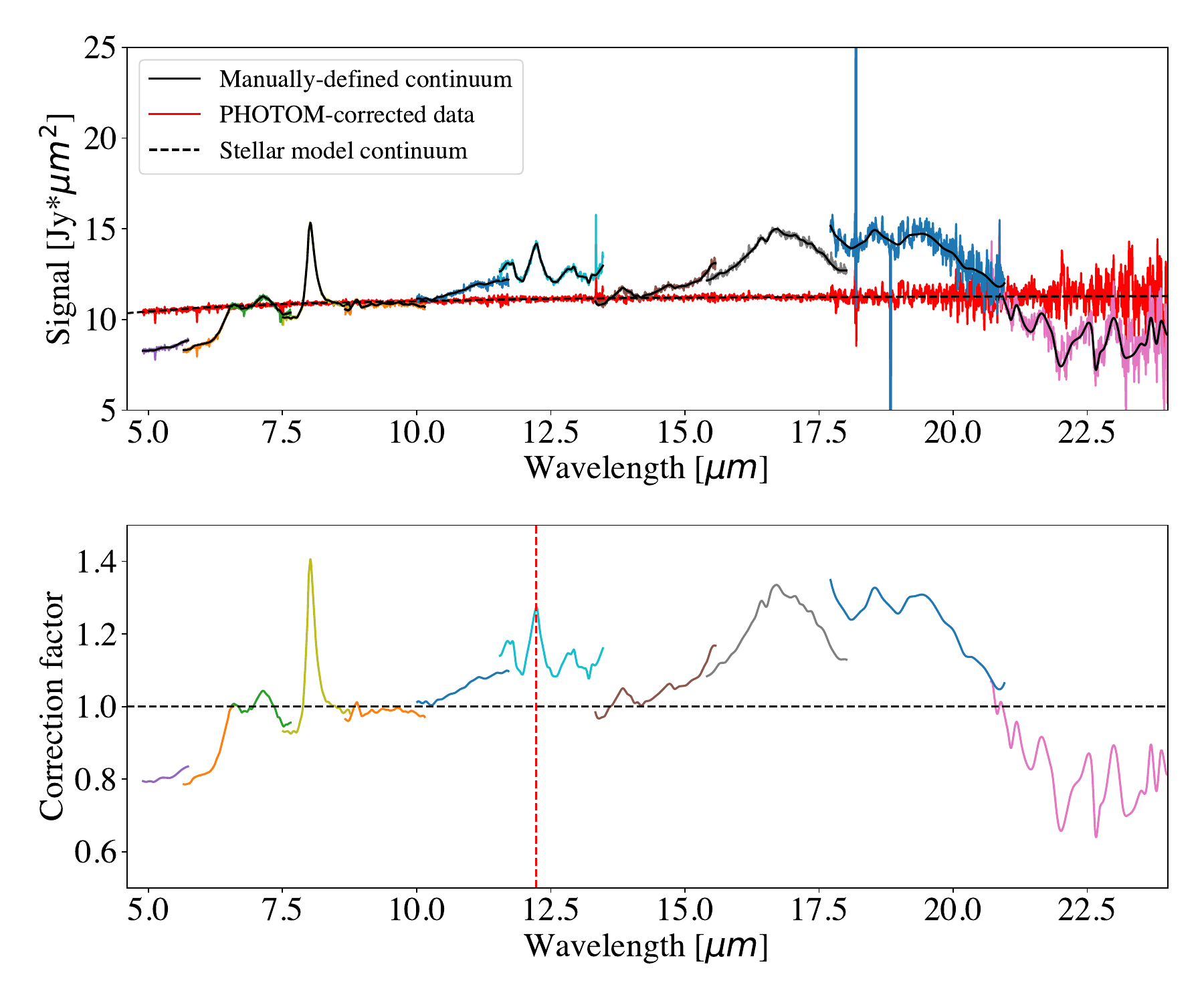}
\caption[]{MRS spectro-photometric correction based on flight data. Top: Colored curves showing the MRS point source integrated spectra in different spectral bands for the A-star HD~163466, which were reduced using the ground-based spectro-photometric solution. The spectrum is cut at 23~$\mu m$ due to the S/N on the source becoming too low beyond that point. A spectral continuum is determined for each spectral band (black solid line). By comparing the manually defined continuum to the theoretical continuum of the A-star, based on \cite{Gordon2022}, a 1D spectro-photometric correction vector is derived for each spectral band. After introducing this correction, the A-star data are reduced once more. The resulting spectrum is shown by the red line. Bottom: One-dimensional spectro-photometric correction vectors. The prominent emission feature at 8~$\mu m$ is an artifact from the MIRI FM ground test campaign. This artifact is corrected in the in-flight spectro-photometric solution. The dotted red line shows the location of the MRS spectral leak at 12.22~$\mu m$ (see text for further details).}
\label{fig:photom_correction}
\end{figure}

\begin{figure}
\centering
\includegraphics[width=0.49\textwidth]{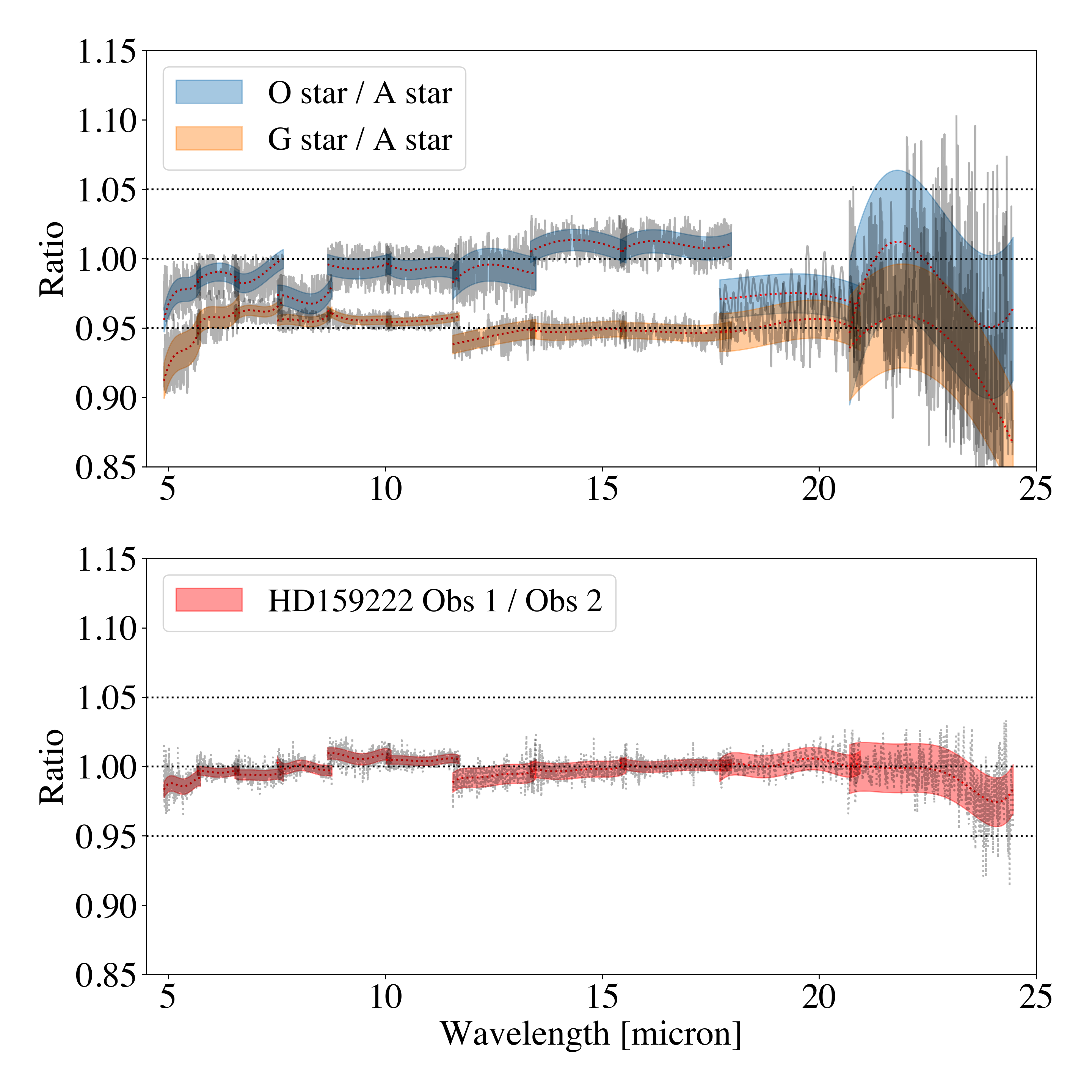}
\caption[]{Repeatability of MRS signal during commissioning. Top: MRS spectro-photometric precision. The presence of the CO molecular band between 4.9-6~$\mu m$ in the G star HD~159222 results in the ratio bending downward in that wavelength range. Bottom: MRS spectro-photometric repeatability evaluated by taking the ratio of the extracted signal of HD~159222, which was observed twice. The measurements were taken 7 days apart. Band 4long is not shown due to the very low S/N in the data.}
\label{fig:photom_precision_and_repeatability}
\end{figure}

Figure~\ref{fig:photom_correction} illustrates the process of correcting the MRS in-flight spectro-photometric solution, where the ground-to-flight correction factor is plotted in the bottom panel. We note the prominent emission feature at 8~$\mu m$, which is an artifact from the MIRI FM test campaign. This artifact is corrected in the in-flight spectro-photometric solution.

\begin{table*}[h!]
\caption{MRS performance values after pipeline processing.}
\label{tab:mrs_performance}
\centering
\begin{tabular}{lccccc}
Spectral & Astrometric & FLT-5 Spectral &  Fringe & Spectro-photometric & Spectro-photometric \\
Band & accuracy [mas] & accuracy [km s$^{-1}$] & contrast [\%] & precision [\%] & repeatability [\%] \\
\hline
1short &  $20 \pm 8$ & $0 \pm 9$& 0.23 & $3.8^{\alpha} \pm 1.3$ & $98.9 \pm 0.6$ \\
1medium & $-20 \pm 10$ & $-3 \pm 10$ & 0.31 & $3.8^{\alpha} \pm 0.4$ & $99.7 \pm 0.3$ \\
1long  & $10 \pm 11$ & $-2 \pm 2$ & 0.30 & $3.8 \pm 1.3$ & $99.5 \pm 0.4$ \\
2short & $40 \pm 7$ & $-1 \pm 7$  & 0.40 & $4.5 \pm 0.4$ & $99.9 \pm 0.4$ \\
2medium & $0 \pm 6$ & $-1 \pm 6$ & 0.40 & $4.4 \pm 0.3$ & $100.8 \pm 0.4\%$ \\
2long & $40 \pm 5$ & $2 \pm 5$ & 0.30 & $4.5 \pm 0.3$ & $100.5 \pm 0.3$ \\
3short & $-20 \pm 7$ & $-3 \pm 5$ & 0.60 & $5.6 \pm 0.7$ & $99.3 \pm 0.7$ \\
3medium & $10 \pm 8$ & $1 \pm 7$ & 0.57 & $5.2 \pm 0.6$ & $99.8 \pm 0.6$ \\
3long & $-20 \pm 9$ & $-13 \pm 27^\beta$ & 0.29 & $5.3 \pm 0.6$ & $100.1 \pm 0.4\%$ \\
4short & $0 \pm 11$ & $-13 \pm 27^\beta$ & 0.53 & $4.9 \pm 1.4$ & $100.2 \pm 0.8\%$ \\
4medium & $0 \pm 15$ & $-13 \pm 27^\beta$ & 1.30 & $5.8 \pm 3.7$ & $99.2 \pm 1.8\%$ \\
4long  & $0 \pm 23$ & $-13 \pm 27^\beta$ & -- & -- & -- \\
\hline
\end{tabular}
\newline
{$^\alpha$}{Precision based on center value of spectral band 1long, which is less impacted by the presence of CO.}\\
{$^\beta$}{Estimated as the mean accuracy of bands 1A-3B in previous wavelength solution FLT-4.}
\end{table*}

At 12.22~$\mu m$ the MRS data show a prominent peak in the signal with a FWHM of 0.1~$\mu m$. This is denoted by the vertical red dotted line in the bottom panel of Fig.~\ref{fig:photom_correction}. The cause of this peak is known; it is present as a systematic excess of flux in the HD~163466 spectrum used to derive the MRS spectro-photometric solution. Specifically, the issue arises from a spectral leakage of the m=2 diffraction grating order being superimposed on the m=1 grating order at 12.22~$\mu m$ (band 3short in Table~\ref{tab:mrs_performance}). Quantitatively, 2.5~\% of the spectral flux density at 6.11~$\mu m$ is added to the 12.22~$\mu m$ wavelengths. This results in a bump in the spectro-photometric response of the 12.22~$\mu m$ range. \footnote{It is important to note that the spectral leak is only partially responsible for the amplitude of the bump around the vertical red dotted line in Fig.~\ref{fig:photom_correction} as the intrinsic correction factor already contains a bump at that location.}. The correction of this effect on the response is described in \citet{gasman22}.

To determine the precision of the derived spectro-photometric solution, we observed the A star HD~163466 (PID~1050), the O star 10~Lac (PID~1524) and the G star HD~159222 (PID~1050). For all three targets, which were observed with the same dither pattern, 3D cubes were built from the observations without applying the spectro-photometric solution. Integrated point source spectra were extracted the same way for the three targets, and the spectra  divided by the stellar model of the continuum in each case \citep{Gordon2022}, resulting in a 1D vector for each band and each target. Using the vectors of HD~163466 as the spectro-photometric calibration reference star, the top panel of Fig.~\ref{fig:photom_precision_and_repeatability} shows the ratio of the 10~Lac and HD~159222 vectors with respect to those of HD~163466. Taking into account the presence of the CO molecular band at the short wavelengths of the G star HD~159222, the maximum deviation recorded across the MRS wavelength range, representing the MRS current spectro-photometric precision, was $5.6 \pm 0.7 \%$ (band 4medium and 4 long are disregarded due to the low S/N). The cause for this deviation is complex, and folds in uncertainties in the data systematics and calibration, as well as uncertainties in the theoretical stellar models used. The deviation will likely decrease as we improve our understanding of the data and the models.

To determine the repeatability of the spectro-photometric calibration, the G star HD~159222 was observed twice over a span of 7 days during commissioning. The bottom panel of Fig.~\ref{fig:photom_precision_and_repeatability} shows the ratio of the extracted signals. The largest deviation is recorded in band 1short where the CO molecular band is located. The likely reason for the deviations across the different bands is linked to a small pointing nonrepeatability discussed in Patapis et al. (in prep.).



\subsection{Straylight}

\begin{figure}
\centering
\includegraphics[width=0.49\textwidth]{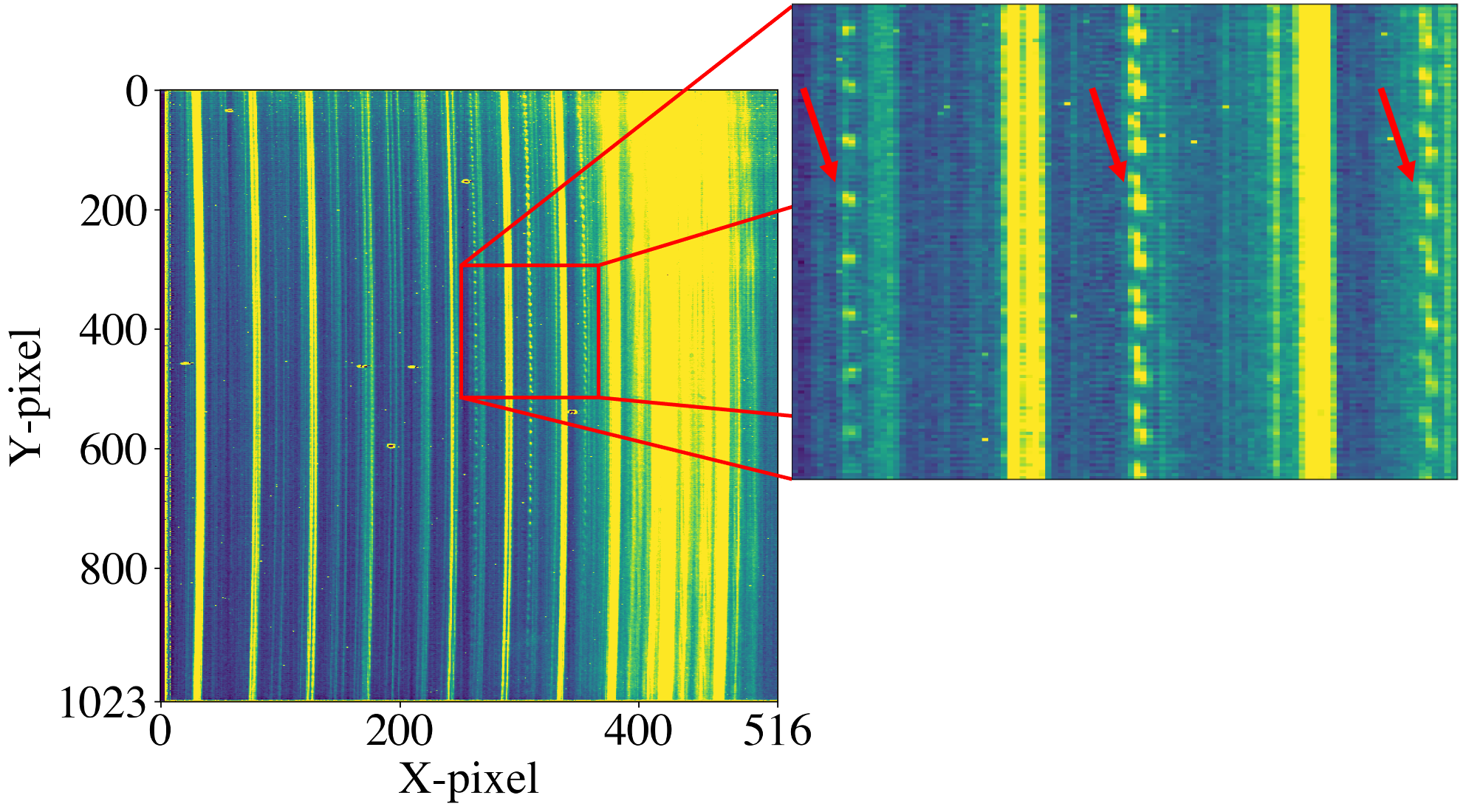}
\caption[]{The "zipper" MRS straylight artifact. When the point source is located on the left side of the MRS image slicer in channel 1, the effect manifests in three slices on the detector (red arrows).}
\label{fig:zipper}
\end{figure}

Light scattering between the MRS optical surfaces, or between an optical surface and structural components can lead to a straylight path to the detectors. The MRS optics were designed with light traps and baffles and coatings to control this effect, but small residual scattering remains. From the detailed analysis of commissioning and Cycle 1 calibration data, only one new source of straylight was detected. Nicknamed the "zipper", this effect is only observed in bright point source observations. The zipper manifests as a series of regular faint emission lines along the detector dispersion direction. Its morphology is shown in Fig.~\ref{fig:zipper}.

Although the origin of the zipper has not been determined at this time, four phenomenological facts have been collected about this effect. (i) The zipper has only been observed in channel 1 of the MRS. (ii) No zipper is visible for dithers that place the point source on the left half of the channel 1 image slicer. (iii) The zipper is sliced by the image slicer (we see it appear in three slices on the detector). (iv) The period of the zipper emission line peaks match that of the respective fringe flats in spectral channel 1 (5 to 8~$\mu m$); however, there appears to be an additional interference, given that every other peak seems to experience destructive interference.

Importantly, the end users of the MRS looking at data cubes built from dithered observations will hardly see the zipper. That is due to the fact that the pipeline drizzles dithered detector frames together, and since the zipper is only present for the dithers with positive along-slice position, the artifact gets averaged-out when combined with the negative along-slice positions where the zipper is absent.

\section{Limitations to MRS sensitivity}
\label{sec:limitations}

\subsection{Dark current versus signal in the dark}

The MRS optics were designed so that the gaps between the slice images on the detector, were large enough to avoid cross-talk between slices, whilst still maximizing the number of detector pixels available for science.  The inter-channel regions were then made somewhat wider, primarily to avoid vignetting, but also to provide regions where the dark current could be monitored. These design choices proved useful in characterizing the scattered light seen in the MRS at short wavelengths. 

One issue that arose very early in the MRS commissioning phase, while monitoring the signal in the wider gap between channels, is the fact that the MRS detector dark current seems to change drastically from one observation to the next Morrison et al. (accepted, PASP). Currently, to address the issue of the varying dark current, the MRS calibration pipeline uses the gap region between the channels to estimate a single pedestal value (mean signal in the dark) which it then subtracts from all the other detector pixels.


\subsection{Cosmic showers}
\begin{figure}[t]
\centering
\includegraphics[width=0.49\textwidth]{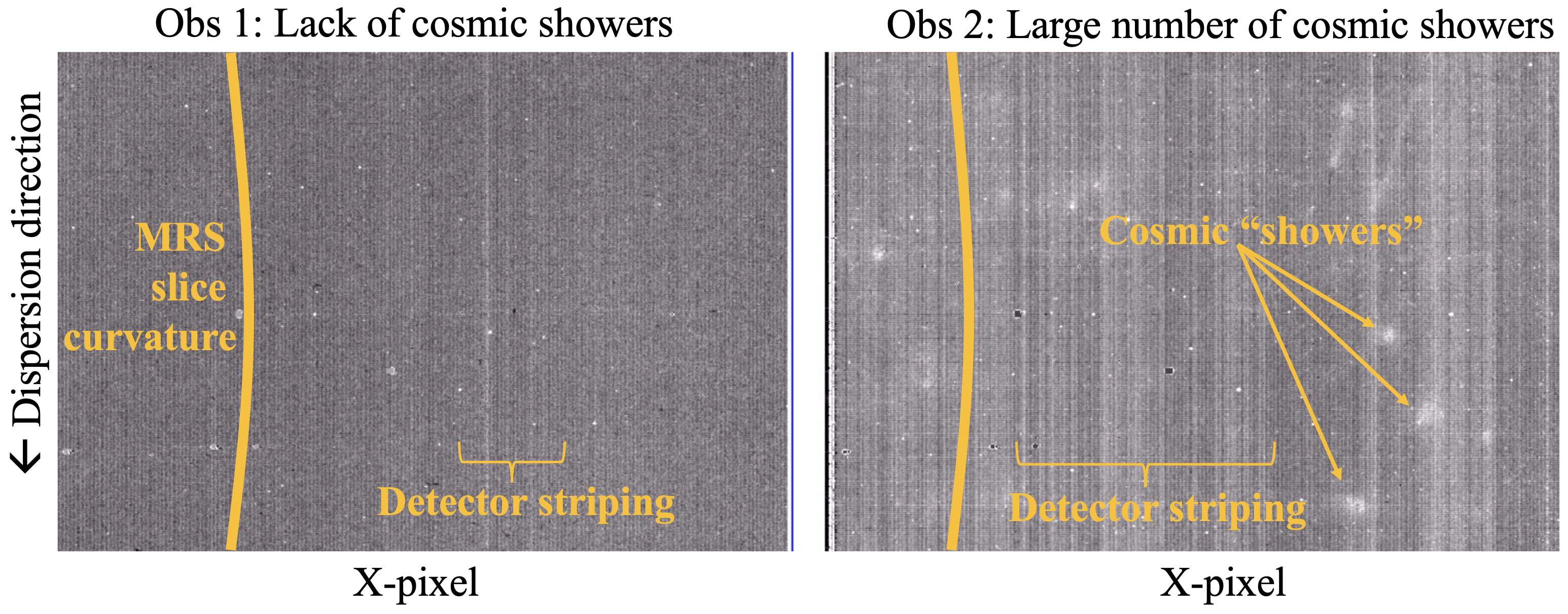}
\caption{Illustration of time-dependence of cosmic showers between MRS observations. Left: MRS detector plane image in one observation that had very few large cosmic ray "shower" events and minimal detector striping from electronic artifacts. Right: Different observation in the same spectral range, with many cosmic showers registered. The signal in the dark is higher than in the left panel, and the detector striping is more prominent. Due to the curvature of the MRS slices dispersed on the detector, the cosmic showers and detector striping map onto the 3D spectral cubes in a nonlinear fashion.}
\label{fig:cosmic_showers}
\end{figure}

Another factor that limits the MRS sensitivity lies in the manifestation of cosmic "showers". Each time the MRS detectors get hit by a single energetic particle (a cosmic ray), the geometric shape of the signal that is recorded on the detector looks like a faint patch. That is to say, the event extends much further out than the original pixel that was affected by the cosmic hit. This formation of secondary cosmic hits is likely caused by the collision of a cosmic ray with a nucleus of aluminum in the MIRI structure causing the observed cosmic "showers" \citep{Pickel2004}. These cosmic showers are illustrated in the right panel of Fig.~\ref{fig:cosmic_showers}. We note that the arrival rate of cosmic rays is variable, with some observations appearing very clean (left panel of Fig.~\ref{fig:cosmic_showers}) while others are full of showers (right panel of Fig.~\ref{fig:cosmic_showers}).

\begin{figure}[h!]
\centering
\includegraphics[width=0.49\textwidth]{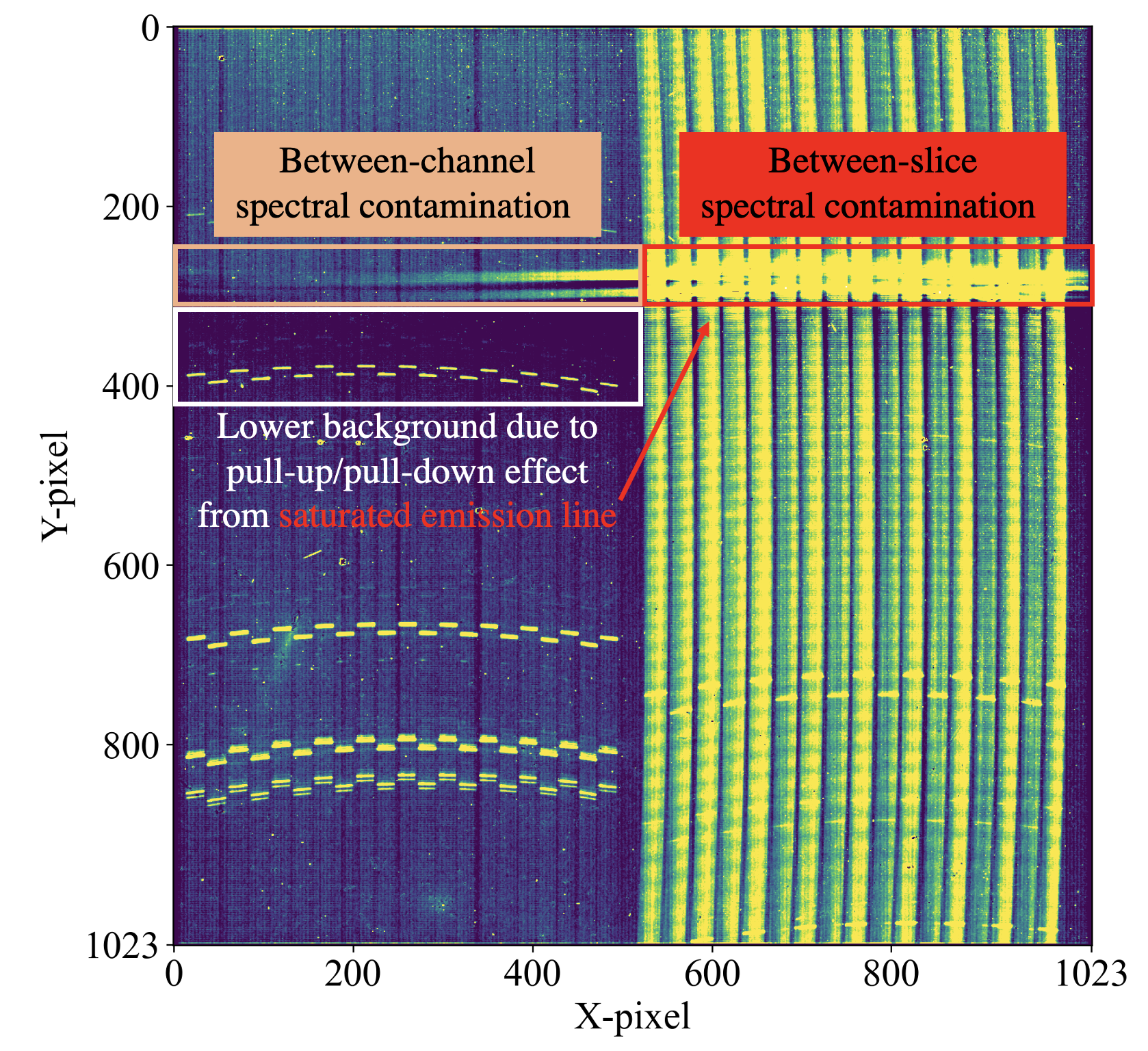}
\caption{Detector image plane of a single exposure on the Cat's Eye Nebula (NGC~6543) showing three kinds of spectral contamination from a bright emission line (Sulfur IV forbidden line at $\lambda = 10.5115$~$\mu m$). The first kind is between dispersed slices, occurring on the right half of the detector. The second kind is between the wavelengths dispersed on the right half of the detector and the slices on the left half of the detector. The third kind of contamination is visible in the detector rows under those impacted by the bright emission line, specifically in the background. This one is lower due to the pull-up/pull-down effect linked to the bright emission line.}
\label{fig:pupd}
\end{figure}

The curvature of the MRS spectra on the detector result in a single cosmic ray shower appearing at multiple, separated locations in the data cube. Detector striping (discussed in Morrison et al., accepted, PASP) impacts the cube signal in a similar way, where brightness differences between detector columns intersect the curved MRS spectral traces.

\subsection{Spectral contamination from bright emission lines}
In Sect.~\ref{subsec:psf_and_scattering} we presented the impact of the MRS detector scattering, and how the PSF in one detector slice contaminates the signal in neighboring slices. An added complexity resulting from the MRS design, is that alternate slices are offset in their wavelength solutions. A pixel recording a wavelength of 5~$\mu m$ in row 390 in one slice, will record the same wavelength of 5~$\mu m$ in row 400 one slice to the right, and in the row 390 to the right again. This alternating pattern is visible in the step profile of the spatially extended emission lines in the left half of the detector in Fig.~\ref{fig:pupd}. As a result, the detector scattering discussed in Sect.~\ref{subsec:psf_and_scattering} is in fact not simply a spatial contamination, but also, simultaneously, a spectral contamination.

There are two other manifestations of spectral contamination that need to be accounted for when studying MRS data. The second type of spectral contamination is from one spectral channel on the detector to the other spectral channel on the same detector. A very bright emission line, such as the one shown spanning the right half of the detector around row number 300 in Fig.~\ref{fig:pupd} (Sulfur IV forbidden line at $\lambda = 10.5115$~$\mu m$), is seen to contaminate the spectrum measured in the left half of the detector. Users studying only the spectral cubes (i.e., not examining the detector image plane) will see a faint emission feature at certain wavelengths in the cube, which are in fact produced by a bright emission line present in a different spectral cube.

\begin{figure}[t]
\centering
\includegraphics[width=0.49\textwidth]{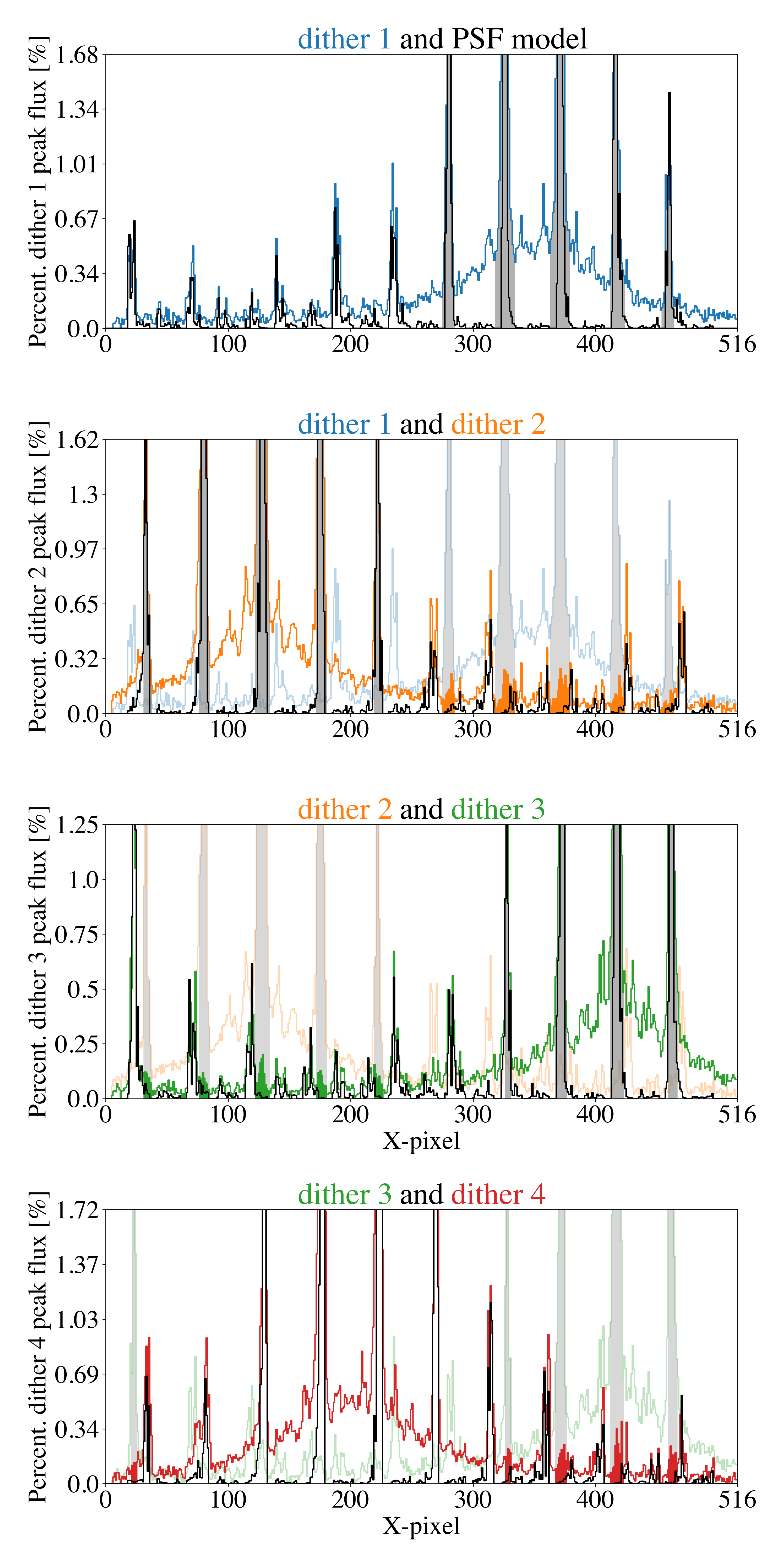}
\caption{MRS persistence detected in point source observation in 4-point dither pattern. First panel: Data of the first dither and the five brightest peaks are filled in gray. A diffraction limited WebbPSF model, which is projected on the detector at the location of the observed point source. Second panel: At the location of the PSF peaks of dither 1, small excesses in flux are recorded in the signal of dither 2 (excess in flux shown by filled data in orange color). Third panel: Persistence caused by dither 2 detector illumination on dither 3 data. Fourth panel: Persistence caused by dither 3 detector illumination on dither 4 data.}
\label{fig:persistence}
\end{figure}

The third type of spectral contamination is caused by the "pull-up/pull-down" electronic cross-talk effect \citep{Dicken2022SPIE}. In Fig.~\ref{fig:pupd} this can be seen on the left half of the detector around row 300, where the detector image goes dark for approximately 100 rows as a result of the bright emission line on the right half of the detector (row 300). The effect can only be addressed using optimal data reduction tools, specifically by using the behavior of the pixels between the spectral channels to bootstrap the correction. In general, we reiterate that there are many advantages to studying (and correcting) the signal in the detector image plane.

\subsection{Persistence}
In the MIRI Imager, the signal from a bright source dissipates to a fraction (<0.01\%) 30 minutes after the source has been removed. For the MRS, we observed a peak level of persistence for point sources between 0.1~\% and 0.2~\% of the peak of the PSF. The appearance of the persistence for the 4-point dithered observation of the G star HD~159222 (PID~1050) is illustrated in Fig.~\ref{fig:persistence}. It is worth noting that the observations of HD~159222 were not saturated, in fact the signal barely reaches half of the detector pixel dynamic range. This suggests that persistence does not only impact very bright targets. 

In the top panel of Fig.~\ref{fig:persistence} we show the data of the first dither of HD~159222 and the five brightest peaks of the PSF are filled in gray. We overplot a diffraction limited WebbPSF model, which we project on the detector at the location of the observed point source. Although the WebbPSF model does not contain the effect of the detector scattering, nor the PSF broadening, nor the subtle details of the real MRS PSF on the detector, it is useful in the context of visually decoupling the persistence from the point source signal. The other three panels in Fig.~\ref{fig:persistence} show subsequent combinations of dithers. 

In the second panel of Fig.~\ref{fig:persistence} we can see that at the location of the PSF peaks of dither 1, small excesses in flux are recorded in the signal of dither 2 (excess in flux shown by filled data in orange color). The PSF model projected at the location of dither 2 strongly suggests that the excess in flux is persistence from the previous point source illumination of dither 1. A similar conclusion is drawn in the third and fourth panel of Fig.~\ref{fig:persistence}. After reconstructing the 3D spectral cubes, the persistence manifests as a faint PSF at the location of the dithers. The correction for the effect in the \textit{JWST} calibration pipeline is part of future calibration work.

\section{Conclusions and future work}
\label{sec:conclusions}

Following the commissioning phase of \textit{JWST} and the MIRI instrument, the MIRI MRS was agreed to have met all of the NASA criteria needed to start its science mission. We have reported on the details of the \textit{JWST} MIRI/MRS performance in flight, including the PSF optical quality and the LSF quality, the astrometric and spectral distortion accuracy, and the spectro-photometric precision. Specifically, we find the following:
\begin{itemize}
    \item The MRS PSF is broadened compared to the ideal diffraction model, and the observed broadening is greatest at short wavelengths. To explain the detected broadening, we identified the speed of the science beam, in conjunction with the ability of the infrared photons to scatter inside the MRS detectors multiple times before being absorbed, as being at the core of the issue.
    \item The overall trend of the MRS LSF matches the expectations from the ground. Some level of uncertainty was introduced due to the fact that the emission lines are resolved.
    \item The astrometric accuracy was gauged relative to the \textit{JWST} global coordinates V2/V3.  
    \item The spectral distortion accuracy was estimated down to below 10~\kms up to band 3medium. The calibration at the longer wavelengths will improve with later calibration cycles.
    \item Spectral fringes, which modulate the spectral baseline of observed astronomical sources, were reduced to below 1.5~\% for all types of source illuminations. The use of flux calibration standards can help mitigate the need for an empirical fringe correction.
    \item The current spectro-photometric precision achieved is 5.6 $\pm$ 0.7~\%.
\end{itemize}

As more data are collected in future JWST calibration cycles, the calibration solutions will continue to improve. Thanks to the effort of the whole MIRI team, we are starting from a very good baseline. Some limitations to the inherent MRS sensitivity -- including the impact of cosmic ray showers, the higher than expected signal in the dark, and the impact of electronic cross-talk -- remain cornerstone issues that will need to be addressed. These will require extensive analysis, and the development of mitigation strategies and algorithms will be important to get the most out of the powerful instrument that is the MIRI/MRS on board \textit{JWST}.

\begin{acknowledgements}
   Ioannis Argyriou, Danny Gasman, Bart Vandenbussche, and Pierre Royer thank the European Space Agency (ESA) and the Belgian Federal Science Policy Office (BELSPO) for their support in the framework of the PRODEX Programme. Ioannis Argyriou would like to personally thank Ruymán Azzollini for all the hard work getting the MRS spectro-photometrically calibrated on the ground, as well as Eliot Malumuth for the ever-insightful discussions about the troublesome fringes. \\
   Patrick Kavanagh thanks the European Space Agency (ESA) and Enterprise Ireland for their support in the framework of the PRODEX Programme. \\
   Javier~Álvarez-Márquez acknowledges support by grant PIB2021-127718NB-100 by the Spanish Ministry of Science and Innovation/State Agency of Research MCIN/AEI/10.13039/501100011033 and by “ERDF A way of making Europe”
   Alvaro Labiano acknowledges the support from Comunidad de Madrid through the Atracción de Talento Investigador Grant 2017-T1/TIC-5213, and PID2019-106280GB-I00 (MCIU/AEI/FEDER,UE). AL also thanks the SBMB crew for their huge effort, great attitude, and incredible professionalism during the intense MRS commissioning campaign, which made possible an absolute Science Verification success. \\ 
   Polychronis Patapis thanks the Swiss Society for Astrophysics and Astronomy (SSAA) for the MERAC Funding and Travel award.\\
   The MRS commissioning team acknowledges the stickies on the wall of the STScI library for reminding us that complex organizational tasks require simple solutions.
   \\
   Jake Harkett thanks STFC for their generous support in funding his Ph.D.\\
   The work presented is the effort of the entire MIRI team and the enthusiasm within the MIRI partnership is a significant factor in its success. MIRI draws on the scientific and technical expertise of the following organizations: Ames Research Center, USA; Airbus Defence and Space, UK; CEA-Irfu, Saclay, France; Centre Spatial de Liége, Belgium; Consejo Superior de Investigaciones Científicas, Spain; Carl Zeiss Optronics, Germany; Chalmers University of Technology, Sweden; Danish Space Research Institute, Denmark; Dublin Institute for Advanced Studies, Ireland; European Space Agency, Netherlands; ETCA, Belgium; ETH Zurich, Switzerland; Goddard Space Flight Center, USA; Institute d'Astrophysique Spatiale, France; Instituto Nacional de Técnica Aeroespacial, Spain; Institute for Astronomy, Edinburgh, UK; Jet Propulsion Laboratory, USA; Laboratoire d'Astrophysique de Marseille (LAM), France; Leiden University, Netherlands; Lockheed Advanced Technology Center (USA); NOVA Opt-IR group at Dwingeloo, Netherlands; Northrop Grumman, USA; Max-Planck Institut für Astronomie (MPIA), Heidelberg, Germany; Laboratoire d’Etudes Spatiales et d'Instrumentation en Astrophysique (LESIA), France; Paul Scherrer Institut, Switzerland; Raytheon Vision Systems, USA; RUAG Aerospace, Switzerland; Rutherford Appleton Laboratory (RAL Space), UK; Space Telescope Science Institute, USA; Toegepast- Natuurwetenschappelijk Onderzoek (TNO-TPD), Netherlands; UK Astronomy Technology Centre, UK; University College London, UK; University of Amsterdam, Netherlands; University of Arizona, USA; University of Bern, Switzerland; University of Cardiff, UK; University of Cologne, Germany; University of Ghent; University of Groningen, Netherlands; University of Leicester, UK; University of Leuven, Belgium; University of Stockholm, Sweden; Utah State University, USA. A portion of this work was carried out at the Jet Propulsion Laboratory, California Institute of Technology, under a contract with the National Aeronautics and Space Administration.

   We would like to thank the following National and International Funding Agencies for their support of the MIRI development: NASA; ESA; Belgian Science Policy Office; Centre Nationale D'Etudes Spatiales (CNES); Danish National Space Centre; Deutsches Zentrum fur Luft-und Raumfahrt (DLR); Enterprise Ireland; Ministerio De Economiá y Competividad; Netherlands Research School for Astronomy (NOVA); Netherlands Organisation for Scientific Research (NWO); Science and Technology Facilities Council; Swiss Space Office; Swedish National Space Board; UK Space Agency.

   We take this opportunity to thank the ESA \textit{JWST} Project team and the NASA Goddard ISIM team for their capable technical support in the development of MIRI, its delivery and successful integration.

\end{acknowledgements}

\bibliography{aanda} 

\begin{thebibliography}{36}
\expandafter\ifx\csname natexlab\endcsname\relax\def\natexlab#1{#1}\fi

\bibitem[{Argyriou(2021)}]{phdthesisYannis}
Argyriou, I. 2021, Calibration of the MIRI instrument on board the James Webb
  Space Telescope,
  \url{https://fys.kuleuven.be/ster/pub/phd-thesis-yannis-argyriou/PhD_Thesis_IoannisArgyriou.pdf}

\bibitem[{{Argyriou} {et~al.}(2020{\natexlab{a}}){Argyriou}, {Rieke},
  {Ressler}, {G{\'a}sp{\'a}r}, \& {Vandenbussche}}]{argyriou2020SPIE}
{Argyriou}, I., {Rieke}, G.~H., {Ressler}, M.~E., {G{\'a}sp{\'a}r}, A., \&
  {Vandenbussche}, B. 2020{\natexlab{a}}, in Society of Photo-Optical
  Instrumentation Engineers (SPIE) Conference Series, Vol. 11454, Society of
  Photo-Optical Instrumentation Engineers (SPIE) Conference Series, 114541P

\bibitem[{{Argyriou} {et~al.}(2020{\natexlab{b}}){Argyriou}, {Wells}, {Glasse},
  {Lee}, {Royer}, {Vandenbussche}, {Malumuth}, {Glauser}, {Kavanagh},
  {Labiano}, {Lahuis}, {Mueller}, \& {Patapis}}]{argyriou2020}
{Argyriou}, I., {Wells}, M., {Glasse}, A., {et~al.} 2020{\natexlab{b}}, \aap,
  641, A150

\bibitem[{{Boccaletti} {et~al.}(2022){Boccaletti}, {Cossou}, {Baudoz},
  {Lagage}, {Dicken}, {Glasse}, {Hines}, {Aguilar}, {Detre}, {Nickson},
  {Noriega-Crespo}, {G{\'a}sp{\'a}r}, {Labiano}, {Stark}, {Rouan}, {Reess},
  {Wright}, {Rieke}, \& {Garcia Marin}}]{coronograph_perf}
{Boccaletti}, A., {Cossou}, C., {Baudoz}, P., {et~al.} 2022, arXiv e-prints,
  arXiv:2207.11080

\bibitem[{Boccaletti {et~al.}(2015)Boccaletti, Lagage, Baudoz, Beichman,
  Bouchet, Cavarroc, Dubreuil, Glasse, Glauser, Hines, Lajoie, Lebreton,
  Perrin, Pueyo, Reess, Rieke, Ronayette, Rouan, Soummer, \&
  Wright}]{miri_pasp_5}
Boccaletti, A., Lagage, P.-O., Baudoz, P., {et~al.} 2015, Publications of the
  Astronomical Society of the Pacific, 127, 633

\bibitem[{{Bohlin} {et~al.}(2014){Bohlin}, {Gordon}, \&
  {Tremblay}}]{Bohlin2014}
{Bohlin}, R.~C., {Gordon}, K.~D., \& {Tremblay}, P.~E. 2014, \pasp, 126, 711

\bibitem[{Bouchet {et~al.}(2015)Bouchet, Garc{\'{\i}}a-Mar{\'{\i}}n, Lagage,
  Amiaux, Augu{\'{e}}res, Bauwens, Blommaert, Chen, Detre, Dicken, Dubreuil,
  Galdemard, Gastaud, Glasse, Gordon, Gougnaud, Guillard, Justtanont, Krause,
  Leboeuf, Longval, Martin, Mazy, Moreau, Olofsson, Ray, Rees, Renotte,
  Ressler, Ronayette, Salasca, Scheithauer, Sykes, Thelen, Wells, Wright, \&
  Wright}]{miri_pasp_3}
Bouchet, P., Garc{\'{\i}}a-Mar{\'{\i}}n, M., Lagage, P.-O., {et~al.} 2015,
  Publications of the Astronomical Society of the Pacific, 127, 612

\bibitem[{{Bryce} {et~al.}(1992){Bryce}, {Meaburn}, {Walsh}, \&
  {Clegg}}]{bryce92}
{Bryce}, M., {Meaburn}, J., {Walsh}, J.~R., \& {Clegg}, R.~E.~S. 1992, \mnras,
  254, 477

\bibitem[{{Dicken} {et~al.}(2022){Dicken}, {Rieke}, {Ressler}, {Morrison},
  {Garcia Marin}, {Argyriou}, {Gordon}, {Regan}, {Cossou}, {Gaspar}, {Glasse},
  {Guillard}, {Labiano}, \& {Wright}}]{Dicken2022SPIE}
{Dicken}, D., {Rieke}, G., {Ressler}, M., {et~al.} 2022, in Society of
  Photo-Optical Instrumentation Engineers (SPIE) Conference Series, Vol. 12180,
  Space Telescopes and Instrumentation 2022: Optical, Infrared, and Millimeter
  Wave, ed. L.~E. {Coyle}, S.~{Matsuura}, \& M.~D. {Perrin}, 121802R

\bibitem[{{Fletcher} {et~al.}(2016){Fletcher}, {Greathouse}, {Orton},
  {Sinclair}, {Giles}, {Irwin}, \& {Encrenaz}}]{fletcher16}
{Fletcher}, L.~N., {Greathouse}, T.~K., {Orton}, G.~S., {et~al.} 2016, \icarus,
  278, 128

\bibitem[{{Fletcher} {et~al.}(2018){Fletcher}, {Orton}, {Sinclair}, {Guerlet},
  {Read}, {Antu{\~n}ano}, {Achterberg}, {Flasar}, {Irwin}, {Bjoraker},
  {Hurley}, {Hesman}, {Segura}, {Gorius}, {Mamoutkine}, \&
  {Calcutt}}]{fletcher18}
{Fletcher}, L.~N., {Orton}, G.~S., {Sinclair}, J.~A., {et~al.} 2018, Nature
  Communications, 9, 3564

\bibitem[{Gasman {et~al.}(2023)Gasman, Argyriou, Sloan, Aringer, Álvarez
  Márquez, Fox, Glasse, Glauser, Jones, Justtanont, Kavanagh, Klaassen,
  Labiano, Larson, Law, Mueller, Nayak, \& Noriega-Crespo}]{gasman22}
Gasman, D., Argyriou, I., Sloan, G., {et~al.} 2023, Astronomy \& Astrophysics,
  673

\bibitem[{{G{\'a}sp{\'a}r} {et~al.}(2021){G{\'a}sp{\'a}r}, {Rieke}, {Guillard},
  {Dicken}, {Gastaud}, {Alberts}, {Morrison}, {Ressler}, {Argyriou}, \&
  {Glasse}}]{Gaspar2021}
{G{\'a}sp{\'a}r}, A., {Rieke}, G.~H., {Guillard}, P., {et~al.} 2021, \pasp,
  133, 014504

\bibitem[{{Glasse} {et~al.}(2015){Glasse}, {Rieke}, {Bauwens},
  {Garc{\'\i}a-Mar{\'\i}n}, {Ressler}, {Rost}, {Tikkanen}, {Vandenbussche}, \&
  {Wright}}]{miri_pasp_9}
{Glasse}, A., {Rieke}, G.~H., {Bauwens}, E., {et~al.} 2015, \pasp, 127, 686

\bibitem[{{Gordon} {et~al.}(2022){Gordon}, {Bohlin}, {Sloan}, {Rieke}, {Volk},
  {Boyer}, {Muzerolle}, {Schlawin}, {Deustua}, {Hines}, {Kraemer}, {Mullally},
  \& {Su}}]{Gordon2022}
{Gordon}, K.~D., {Bohlin}, R., {Sloan}, G.~C., {et~al.} 2022, \aj, 163, 267

\bibitem[{{Hickson} {et~al.}(1992){Hickson}, {Mendes de Oliveira}, {Huchra}, \&
  {Palumbo}}]{hickson92}
{Hickson}, P., {Mendes de Oliveira}, C., {Huchra}, J.~P., \& {Palumbo}, G.~G.
  1992, \apj, 399, 353

\bibitem[{{Irwin} {et~al.}(2008){Irwin}, {Teanby}, {de Kok}, {Fletcher},
  {Howett}, {Tsang}, {Wilson}, {Calcutt}, {Nixon}, \& {Parrish}}]{Irwin2008}
{Irwin}, P.~G.~J., {Teanby}, N.~A., {de Kok}, R., {et~al.} 2008, \jqsrt, 109,
  1136

\bibitem[{{Jones} {et~al.}(2023){Jones}, {{\'A}lvarez-M{\'a}rquez}, {Sloan},
  {Kavanagh}, {Argyriou}, {Law}, {Labiano}, {Patapis}, {Mueller}, {Larson},
  {Bright}, {Klaassen}, {Fox}, {Gasman}, {Geers}, {Glauser}, {Guillard},
  {Nayak}, {Noriega-Crespo}, {Ressler}, {Sargent}, {Temim}, {Vandenbussche}, \&
  {Garc{\'\i}a Mar{\'\i}n}}]{jones23}
{Jones}, O.~C., {{\'A}lvarez-M{\'a}rquez}, J., {Sloan}, G.~C., {et~al.} 2023,
  \mnras [\eprint[arXiv]{2301.13233}]

\bibitem[{Kendrew {et~al.}(2015)Kendrew, Scheithauer, Bouchet, Amiaux,
  Azzollini, Bouwman, Chen, Dubreuil, Fischer, Glasse, Greene, Lagage, Lahuis,
  Ronayette, Wright, \& Wright}]{miri_pasp_4}
Kendrew, S., Scheithauer, S., Bouchet, P., {et~al.} 2015, Publications of the
  Astronomical Society of the Pacific, 127, 623

\bibitem[{{Kester} {et~al.}(2003){Kester}, {Beintema}, \&
  {Lutz}}]{sws_fringes_and_models}
{Kester}, D.~J.~M., {Beintema}, D.~A., \& {Lutz}, D. 2003, in ESA Special
  Publication, Vol. 481, The Calibration Legacy of the ISO Mission, ed.
  L.~{Metcalfe}, A.~{Salama}, S.~B. {Peschke}, \& M.~F. {Kessler}, 375

\bibitem[{{Labiano} {et~al.}(2021){Labiano}, {Argyriou},
  {{\'A}lvarez-M{\'a}rquez}, {Glasse}, {Glauser}, {Patapis}, {Law}, {Brandl},
  {Justtanont}, {Lahuis}, {Mart{\'\i}nez-Galarza}, {Mueller}, {Noriega-Crespo},
  {Royer}, {Shaughnessy}, \& {Vandenbussche}}]{labiano2021}
{Labiano}, A., {Argyriou}, I., {{\'A}lvarez-M{\'a}rquez}, J., {et~al.} 2021,
  \aap, 656, A57

\bibitem[{{Labiano} {et~al.}(2016){Labiano}, {Azzollini}, {Bailey}, {Beard},
  {Dicken}, {Garc{\'\i}a-Mar{\'\i}n}, {Geers}, {Glasse}, {Glauser}, {Gordon},
  {Justtanont}, {Klaassen}, {Lahuis}, {Law}, {Morrison}, {M{\"u}ller}, {Rieke},
  {Vandenbussche}, \& {Wright}}]{labiano2016}
{Labiano}, A., {Azzollini}, R., {Bailey}, J., {et~al.} 2016, in Society of
  Photo-Optical Instrumentation Engineers (SPIE) Conference Series, Vol. 9910,
  Observatory Operations: Strategies, Processes, and Systems VI, ed. A.~B.
  {Peck}, R.~L. {Seaman}, \& C.~R. {Benn}, 99102W

\bibitem[{{Lahuis} \& {Boogert}(2003)}]{fringes_sirtf_irs}
{Lahuis}, F. \& {Boogert}, A. 2003, in SFChem 2002: Chemistry as a Diagnostic
  of Star Formation, ed. C.~L. {Curry} \& M.~{Fich}, 335

\bibitem[{{Li Causi} {et~al.}(2016){Li Causi}, {Lee}, {Vitali}, {Royer}, \&
  {Oliva}}]{moons_defocus}
{Li Causi}, G., {Lee}, D., {Vitali}, F., {Royer}, F., \& {Oliva}, E. 2016, in
  Society of Photo-Optical Instrumentation Engineers (SPIE) Conference Series,
  Vol. 9908, Ground-based and Airborne Instrumentation for Astronomy VI, ed.
  C.~J. {Evans}, L.~{Simard}, \& H.~{Takami}, 99088P

\bibitem[{{Malumuth} {et~al.}(2003){Malumuth}, {Hill}, {Gull}, {Woodgate},
  {Bowers}, {Kimble}, {Lindler}, {Plait}, \& {Blouke}}]{stis_fringing_malumuth}
{Malumuth}, E.~M., {Hill}, R.~S., {Gull}, T., {et~al.} 2003, \pasp, 115, 218

\bibitem[{{O'Connor} {et~al.}(2006){O'Connor}, {Radeka}, {Figer}, {Geary},
  {Gilmore}, {Oliver}, {Stubbs}, {Takacs}, \& {Tyson}}]{lsst_defocus}
{O'Connor}, P., {Radeka}, V., {Figer}, D., {et~al.} 2006, in Society of
  Photo-Optical Instrumentation Engineers (SPIE) Conference Series, Vol. 6276,
  Society of Photo-Optical Instrumentation Engineers (SPIE) Conference Series,
  ed. D.~A. {Dorn} \& A.~D. {Holland}, 62761W

\bibitem[{{Perrin} {et~al.}(2014){Perrin}, {Sivaramakrishnan}, {Lajoie},
  {Elliott}, {Pueyo}, {Ravindranath}, \& {Albert}}]{Perrin2014}
{Perrin}, M.~D., {Sivaramakrishnan}, A., {Lajoie}, C.-P., {et~al.} 2014, in
  Society of Photo-Optical Instrumentation Engineers (SPIE) Conference Series,
  Vol. 9143, Space Telescopes and Instrumentation 2014: Optical, Infrared, and
  Millimeter Wave, ed. J.~{Oschmann}, Jacobus~M., M.~{Clampin}, G.~G. {Fazio},
  \& H.~A. {MacEwen}, 91433X

\bibitem[{{Perrin} {et~al.}(2012){Perrin}, {Soummer}, {Elliott}, {Lallo}, \&
  {Sivaramakrishnan}}]{Perrin2012}
{Perrin}, M.~D., {Soummer}, R., {Elliott}, E.~M., {Lallo}, M.~D., \&
  {Sivaramakrishnan}, A. 2012, in Society of Photo-Optical Instrumentation
  Engineers (SPIE) Conference Series, Vol. 8442, Space Telescopes and
  Instrumentation 2012: Optical, Infrared, and Millimeter Wave, ed. M.~C.
  {Clampin}, G.~G. {Fazio}, H.~A. {MacEwen}, \& J.~{Oschmann}, Jacobus~M.,
  84423D

\bibitem[{{Pickel} {et~al.}(2004){Pickel}, {Reed}, {Marshall}, {Waczynski},
  {Polidan}, {Johnson}, {McMurray}, {McKelvey}, {Ennico}, {Johnson}, \&
  {Gee}}]{Pickel2004}
{Pickel}, J.~C., {Reed}, R.~A., {Marshall}, P.~W., {et~al.} 2004, in Society of
  Photo-Optical Instrumentation Engineers (SPIE) Conference Series, Vol. 5487,
  Optical, Infrared, and Millimeter Space Telescopes, ed. J.~C. {Mather},
  698--709

\bibitem[{{Reid} \& {Parker}(2006)}]{ReidParker2006}
{Reid}, W.~A. \& {Parker}, Q.~A. 2006, \mnras, 373, 521

\bibitem[{{Ressler} {et~al.}(2015){Ressler}, {Sukhatme}, {Franklin}, {Mahoney},
  {Thelen}, {Bouchet}, {Colbert}, {Cracraft}, {Dicken}, {Gastaud}, {Goodson},
  {Eccleston}, {Moreau}, {Rieke}, \& {Schneider}}]{miri_pasp_8}
{Ressler}, M.~E., {Sukhatme}, K.~G., {Franklin}, B.~R., {et~al.} 2015, \pasp,
  127, 675

\bibitem[{Rieke {et~al.}(2015)Rieke, Ressler, Morrison, Bergeron, Bouchet,
  Garc{\'{\i}}a-Mar{\'{\i}}n, Greene, Regan, Sukhatme, \& Walker}]{miri_pasp_7}
Rieke, G.~H., Ressler, M.~E., Morrison, J.~E., {et~al.} 2015, Publications of
  the Astronomical Society of the Pacific, 127, 665

\bibitem[{{Rigby} {et~al.}(2023){Rigby}, {Perrin}, {McElwain}, {Kimble},
  {Friedman}, {Lallo}, {Doyon}, {Feinberg}, {Ferruit}, {Glasse}, {Rieke},
  {Rieke}, {Wright}, {Willott}, {Colon}, {Milam}, {Neff}, {Stark}, {Valenti},
  {Abell}, {Abney}, {Abul-Huda}, {Scott Acton}, {Adams}, {Adler}, {Aguilar},
  {Ahmed}, {Albert}, {Alberts}, {Aldridge}, {Allen}, {Altenburg},
  {{\'A}lvarez-M{\'a}rquez}, {Alves de Oliveira}, {Andersen}, {Anderson},
  {Anderson}, {Argyriou}, {Armstrong}, {Arribas}, {Artigau}, {Arvai},
  {Atkinson}, {Bacon}, {Bair}, {Banks}, {Barrientes}, {Barringer}, {Bartosik},
  {Bast}, {Baudoz}, {Beatty}, {Bechtold}, {Beck}, {Bergeron}, {Bergkoetter},
  {Bhatawdekar}, {Birkmann}, {Blazek}, {Blome}, {Boccaletti}, {B{\"o}ker},
  {Boia}, {Bonaventura}, {Bond}, {Bosley}, {Boucarut}, {Bourque}, {Bouwman},
  {Bower}, {Bowers}, {Boyer}, {Bradley}, {Brady}, {Braun}, {Breda},
  {Bresnahan}, {Bright}, {Britt}, {Bromenschenkel}, {Brooks}, {Brooks},
  {Brown}, {Brown}, {Brown}, {Bunker}, {Burger}, {Bushouse}, {Cale}, {Cameron},
  {Cameron}, {Canipe}, {Caplinger}, {Caputo}, {Cara}, {Carey}, {Carniani},
  {Carrasquilla}, {Carruthers}, {Case}, {Catherine}, {Chance}, {Chapman},
  {Charlot}, {Charlow}, {Chayer}, {Chen}, {Cherinka}, {Chichester}, {Chilton},
  {Chonis}, {Clampin}, {Clark}, {Clark}, {Coe}, {Coleman}, {Comber}, {Comeau},
  {Connolly}, {Cooper}, {Cooper}, {Coppock}, {Correnti}, {Cossou}, {Coulais},
  {Coyle}, {Cracraft}, {Curti}, {Cuturic}, {Davis}, {Davis}, {Dean}, {DeLisa},
  {deMeester}, {Dencheva}, {Dencheva}, {DePasquale}, {Deschenes}, {Hunor
  Detre}, {Diaz}, {Dicken}, {DiFelice}, {Dillman}, {Dixon}, {Doggett},
  {Donaldson}, {Douglas}, {DuPrie}, {Dupuis}, {Durning}, {Easmin}, {Eck},
  {Edeani}, {Egami}, {Ehrenwinkler}, {Eisenhamer}, {Eisenhower}, {Elie},
  {Elliott}, {Elliott}, {Ellis}, {Engesser}, {Espinoza}, {Etienne}, {Etxaluze},
  {Falini}, {Feeney}, {Ferry}, {Filippazzo}, {Fincham}, {Fix}, {Flagey},
  {Florian}, {Flynn}, {Fontanella}, {Ford}, {Forshay}, {Fox}, {Franz}, {Fu},
  {Fullerton}, {Galkin}, {Galyer}, {Garc{\'\i}a Mar{\'\i}n}, {Gardner},
  {Gardner}, {Garland}, {Garrett}, {Gasman}, {Gaspar}, {Gaudreau}, {Gauthier},
  {Geers}, {Geithner}, {Gennaro}, {Giardino}, {Girard}, {Giuliano},
  {Glassmire}, {Glauser}, {Glazer}, {Godfrey}, {Golimowski}, {Gollnitz},
  {Gong}, {Gonzaga}, {Gordon}, {Gordon}, {Goudfrooij}, {Greene}, {Greenhouse},
  {Grimaldi}, {Groebner}, {Grundy}, {Guillard}, {Gutman}, {Ha}, {Haderlein},
  {Hagedorn}, {Hainline}, {Haley}, {Hami}, {Hamilton}, {Hammel}, {Hansen},
  {Harkins}, {Harr}, {Hart}, {Hart}, {Hartig}, {Hashimoto}, {Haskins},
  {Hathaway}, {Havey}, {Hayden}, {Hecht}, {Heller-Boyer}, {Henriques}, {Henry},
  {Hermann}, {Hernandez}, {Hesman}, {Hicks}, {Hilbert}, {Hines}, {Hoffman},
  {Holfeltz}, {Holler}, {Hoppa}, {Hott}, {Howard}, {Howard}, {Hunter},
  {Hunter}, {Hurst}, {Husemann}, {Hustak}, {Ilinca Ignat}, {Illingworth},
  {Irish}, {Jackson}, {Jahromi}, {Jakobsen}, {James}, {James}, {Januszewski},
  {Jenkins}, {Jirdeh}, {Johnson}, {Johnson}, {Jones}, {Jones}, {Jones},
  {Jones}, {Jordan}, {Jordan}, {Jurczyk}, {Jurling}, {Kaleida}, {Kalmanson},
  {Kammerer}, {Kang}, {Kao}, {Karakla}, {Kavanagh}, {Kelly}, {Kendrew},
  {Kennedy}, {Kenny}, {Keski-kuha}, {Keyes}, {Kidwell}, {Kinzel}, {Kirk},
  {Kirkpatrick}, {Kirshenblat}, {Klaassen}, {Knapp}, {Scott Knight},
  {Knollenberg}, {Koehler}, {Koekemoer}, {Kovacs}, {Kulp}, {Kumari},
  {Kyprianou}, {La Massa}, {Labador}, {Labiano}, {Lagage}, {Lajoie}, {Lallo},
  {Lam}, {Lamb}, {Lambros}, {Lampenfield}, {Langston}, {Larson}, {Law},
  {Lawrence}, {Lee}, {Leisenring}, {Lepo}, {Leveille}, {Levenson}, {Levine},
  {Levy}, {Lewis}, {Lewis}, {Libralato}, {Lightsey}, {Link}, {Liu}, {Lo},
  {Lockwood}, {Logue}, {Long}, {Long}, {Loomis}, {Lopez-Caniego}, {Lorenzo
  Alvarez}, {Love-Pruitt}, {Lucy}, {Luetzgendorf}, {Maghami}, {Maiolino},
  {Major}, {Malla}, {Malumuth}, {Manjavacas}, {Mannfolk}, {Marrione},
  {Marston}, {Martel}, {Maschmann}, {Masci}, {Masciarelli}, {Maszkiewicz},
  {Mather}, {McKenzie}, {McLean}, {McMaster}, {Melbourne}, {Mel{\'e}ndez},
  {Menzel}, {Merz}, {Meyett}, {Meza}, {Miskey}, {Misselt}, {Moller},
  {Morrison}, {Morse}, {Moseley}, {Mosier}, {Mountain}, {Mueckay}, {Mueller},
  {Mullally}, {Murphy}, {Murray}, {Murray}, {Mustelier}, {Muzerolle},
  {Mycroft}, {Myers}, {Myrick}, {Nanavati}, {Nance}, {Nayak}, {Naylor},
  {Nelan}, {Nickson}, {Nielson}, {Nieto-Santisteban}, {Nikolov},
  {Noriega-Crespo}, {O'Shaughnessy}, {O'Sullivan}, {Ochs}, {Ogle}, {Oleszczuk},
  {Olmsted}, {Osborne}, {Ottens}, {Owens}, {Pacifici}, {Pagan}, {Page}, {Park},
  {Parrish}, {Patapis}, {Paul}, {Pauly}, {Pavlovsky}, {Pedder}, {Peek},
  {Pena-Guerrero}, {Penanen}, {Perez}, {Perna}, {Perriello}, {Phillips},
  {Pietraszkiewicz}, {Pinaud}, {Pirzkal}, {Pitman}, {Piwowar}, {Platais},
  {Player}, {Plesha}, {Pollizi}, {Polster}, {Pontoppidan}, {Porterfield},
  {Proffitt}, {Pueyo}, {Pulliam}, {Quirt}, {Quispe Neira}, {Ramos Alarcon},
  {Ramsay}, {Rapp}, {Rapp}, {Rauscher}, {Ravindranath}, {Rawle}, {Regan},
  {Reichard}, {Reis}, {Ressler}, {Rest}, {Reynolds}, {Rhue}, {Richon},
  {Rickman}, {Ridgaway}, {Ritchie}, {Rix}, {Robberto}, {Robinson}, {Robinson},
  {Robinson}, {Rock}, {Rodriguez}, {Rodriguez Del Pino}, {Roellig}, {Rohrbach},
  {Roman}, {Romelfanger}, {Rose}, {Roteliuk}, {Roth}, {Rothwell}, {Rowlands},
  {Roy}, {Royer}, {Royle}, {Rui}, {Rumler}, {Runnels}, {Russ}, {Rustamkulov},
  {Ryden}, {Ryer}, {Sabata}, {Sabatke}, {Sabbi}, {Samuelson}, {Sapp},
  {Sappington}, {Sargent}, {Sauer}, {Scheithauer}, {Schlawin}, {Schlitz},
  {Schmitz}, {Schneider}, {Schreiber}, {Schulze}, {Schwab}, {Scott}, {Sembach},
  {Shanahan}, {Shaughnessy}, {Shaw}, {Shawger}, {Shay}, {Sheehan}, {Shen},
  {Sherman}, {Shiao}, {Shih}, {Shivaei}, {Sienkiewicz}, {Sing}, {Sirianni},
  {Sivaramakrishnan}, {Skipper}, {Sloan}, {Slocum}, {Slowinski}, {Smith},
  {Smith}, {Smith}, {Smith}, {Snyder}, {Soh}, {Tony Sohn}, {Soto}, {Spencer},
  {Stallcup}, {Stansberry}, {Starr}, {Starr}, {Stewart}, {Stiavelli},
  {Straughn}, {Strickland}, {Stys}, {Summers}, {Sun}, {Sunnquist}, {Swade},
  {Swam}, {Swaters}, {Swoish}, {Taylor}, {Taylor}, {Te Plate}, {Tea}, {Teague},
  {Telfer}, {Temim}, {Thatte}, {Thompson}, {Thompson}, {Thomson}, {Tikkanen},
  {Tippet}, {Todd}, {Toolan}, {Tran}, {Trejo}, {Truong}, {Tsukamoto},
  {Tustain}, {Tyra}, {Ubeda}, {Underwood}, {Uzzo}, {Van Campen}, {Vandal},
  {Vandenbussche}, {Vila}, {Volk}, {Wahlgren}, {Waldman}, {Walker}, {Wander},
  {Warfield}, {Warner}, {Wasiak}, {Watkins}, {Weaver}, {Weilert}, {Weiser},
  {Weiss}, {Weissman}, {Welty}, {West}, {Wheate}, {Wheatley}, {Wheeler},
  {White}, {Whiteaker}, {Whitehouse}, {Whiteleather}, {Whitman}, {Williams},
  {Willmer}, {Willoughby}, {Wilson}, {Wirth}, {Wislowski}, {Wolf}, {Wolfe},
  {Wolff}, {Workman}, {Wright}, {Wu}, {Wu}, {Wymer}, {Yates}, {Yeager},
  {Yeates}, {Yerger}, {Yoon}, {Young}, {Yu}, {Zak}, {Zeidler}, {Zhou},
  {Zielinski}, {Zincke}, \& {Zonak}}]{rigby23}
{Rigby}, J., {Perrin}, M., {McElwain}, M., {et~al.} 2023, \pasp, 135, 048001

\bibitem[{Wells {et~al.}(2015)Wells, Pel, Glasse, Wright, Aitink-Kroes,
  Azzollini, Beard, Brandl, Gallie, Geers, \& et~al.}]{Wells_2015}
Wells, M., Pel, J.-W., Glasse, A., {et~al.} 2015, Publications of the
  Astronomical Society of the Pacific, 127, 646–664

\bibitem[{{Wright} {et~al.}(2023){Wright}, {Rieke}, {Glasse}, {Ressler},
  {Garc{\'\i}a Mar{\'\i}n}, {Aguilar}, {Alberts}, {{\'A}lvarez-M{\'a}rquez},
  {Argyriou}, {Banks}, {Baudoz}, {Boccaletti}, {Bouchet}, {Bouwman}, {Brandl},
  {Breda}, {Bright}, {Cale}, {Colina}, {Cossou}, {Coulais}, {Cracraft}, {De
  Meester}, {Dicken}, {Engesser}, {Etxaluze}, {Fox}, {Friedman}, {Fu},
  {Gasman}, {G{\'a}sp{\'a}r}, {Gastaud}, {Geers}, {Glauser}, {Gordon},
  {Greene}, {Greve}, {Grundy}, {G{\"u}del}, {Guillard}, {Haderlein},
  {Hashimoto}, {Henning}, {Hines}, {Holler}, {Detre}, {Jahromi}, {James},
  {Jones}, {Justtanont}, {Kavanagh}, {Kendrew}, {Klaassen}, {Krause},
  {Labiano}, {Lagage}, {Lambros}, {Larson}, {Law}, {Lee}, {Libralato},
  {Alverez}, {Meixner}, {Morrison}, {Mueller}, {Murray}, {Mycroft}, {Myers},
  {Nayak}, {Naylor}, {Nickson}, {Noriega-Crespo}, {{\"O}stlin}, {O'Sullivan},
  {Ottens}, {Patapis}, {Penanen}, {Pietraszkiewicz}, {Ray}, {Regan},
  {Roteliuk}, {Royer}, {Samara-Ratna}, {Samuelson}, {Sargent}, {Scheithauer},
  {Schneider}, {Schreiber}, {Shaughnessy}, {Sheehan}, {Shivaei}, {Sloan},
  {Tamas}, {Teague}, {Temim}, {Tikkanen}, {Tustain}, {van Dishoeck},
  {Vandenbussche}, {Weilert}, {Whitehouse}, \& {Wolff}}]{wright23}
{Wright}, G.~S., {Rieke}, G.~H., {Glasse}, A., {et~al.} 2023, \pasp, 135,
  048003

\bibitem[{Wright {et~al.}(2015)Wright, Wright, Goodson, Rieke, Aitink-Kroes,
  Amiaux, Aricha-Yanguas, Azzollini, Banks, Barrado-Navascues,
  Belenguer-Davila, Bloemmart, Bouchet, Brandl, Colina, Örs Detre,
  Diaz-Catala, Eccleston, Friedman, Garc{\'{\i}}a-Mar{\'{\i}}n, Güdel, Glasse,
  Glauser, Greene, Groezinger, Grundy, Hastings, Henning, Hofferbert, Hunter,
  Jessen, Justtanont, Karnik, Khorrami, Krause, Labiano, Lagage, Langer, Lemke,
  Lim, Lorenzo-Alvarez, Mazy, McGowan, Meixner, Morris, Morrison, Müller,
  rgaard Nielson, Olofsson, O'Sullivan, Pel, Penanen, Petach, Pye, Ray,
  Renotte, Renouf, Ressler, Samara-Ratna, Scheithauer, Schneider, Shaughnessy,
  Stevenson, Sukhatme, Swinyard, Sykes, Thatcher, Tikkanen, van Dishoeck,
  Waelkens, Walker, Wells, \& Zhender}]{miri_pasp_2}
Wright, G.~S., Wright, D., Goodson, G.~B., {et~al.} 2015, Publications of the
  Astronomical Society of the Pacific, 127, 595

\end{thebibliography}
\bibliographystyle{aa} 

\end{document}